\documentclass{article}
\PassOptionsToPackage{numbers, compress}{natbib}
\usepackage[preprint]{neurips_2025}
\usepackage[utf8]{inputenc} % allow utf-8 input
\usepackage[T1]{fontenc}    % use 8-bit T1 fonts
\usepackage{hyperref}       % hyperlinks
\usepackage{url}            % simple URL typesetting
\usepackage{booktabs}       % professional-quality tables
\usepackage{amsfonts}       % blackboard math symbols
\usepackage{nicefrac}       % compact symbols for 1/2, etc.
\usepackage{microtype}      % microtypography
\usepackage[dvipsnames]{xcolor}         % colors
\usepackage{float} % allow 'h' float specifier
\usepackage{xcolor}
\usepackage{multirow}
\usepackage{float} 
\usepackage{latexsym}
\usepackage{graphicx}
\usepackage{amsmath}
\newcommand{\methodName}{\texttt{ShiftySpeech}}
\usepackage{colortbl} 
\usepackage{wrapfig}
\usepackage{makecell}
\usepackage{enumitem}
\usepackage{adjustbox}

\usepackage{titlesec}
\definecolor{darkgreen}{rgb}{0.0, 0.5, 0.0}

 \usepackage{amssymb}
 \usepackage{amsmath}
\usepackage{amssymb}
\usepackage{pifont}
\usepackage{xcolor}
\usepackage{booktabs}
\usepackage{graphicx}
\usepackage{comment}
\usepackage{nameref}
\newcommand{\cmark}{\textcolor{green!60!black}{\ding{51}}} 
\newcommand{\xmark}{\textcolor{red}{\ding{55}}}    

\def\equationautorefname~#1\null{(#1)\null}
\def\itemautorefname~#1\null{(#1)\null}
\def\sectionautorefname~#1\null{\S#1\null}
\def\subsectionautorefname~#1\null{\S#1\null}
\def\subsubsectionautorefname~#1\null{\S#1\null}

\title{ShiftySpeech: A Large-Scale Synthetic Speech Dataset with Distribution Shifts}
 
\usepackage[textwidth=\textwidth,textsize=small]{todonotes}
\author{%
Ashi Garg$^{1}$ \quad Zexin Cai$^{1}$ \quad Lin Zhang$^1$  \quad
\textbf{Henry Li Xinyuan}$^1$  \\ \textbf{Leibny Paola Garc\'ia-Perera}$^{1}$ \quad \textbf{Kevin Duh}$^{1}$ \quad 
\textbf{Sanjeev Khudanpur}$^{1}$  \quad \textbf{Matthew Wiesner}$^{1}$ \\ \textbf{Nicholas Andrews}$^{1}$ \\ \\
$^1$Johns Hopkins University
}
\begin{document}

\maketitle

\begin{abstract}\label{abstract}
The problem of synthetic speech detection has enjoyed considerable attention, with recent methods achieving low error rates across several established benchmarks. However, to what extent can low error rates on academic benchmarks translate to more realistic conditions? 
In practice, while the training set is fixed at one point in time, test-time conditions may exhibit distribution shifts relative to the training conditions, such as changes in speaker characteristics, emotional expressiveness, language and acoustic conditions, and the emergence of novel synthesis methods. 
Although some existing datasets target subsets of these distribution shifts, systematic analysis remains difficult due to inconsistencies between source data and synthesis systems across datasets. This difficulty is further exacerbated by the rapid development of new text-to-speech (TTS) and vocoder systems, which continually expand the diversity of synthetic speech. To enable systematic benchmarking of model performance under distribution shifts, we introduce \methodName, a large-scale benchmark comprising over 3,000 hours of synthetic speech across 7 source domains, 6 TTS systems, 12 vocoders, and 3 languages. \methodName  \ is specifically designed to evaluate model generalization under controlled distribution shifts while ensuring broad coverage of modern synthetic speech generation techniques. It fills a key gap in current benchmarks by supporting fine-grained, controlled analysis of generalization robustness. All tested distribution shifts significantly degrade detection performance of state-of-the-art detection approaches based on self-supervised features. Overall, our findings suggest that reliance on synthetic speech detection methods in production environments should be carefully evaluated based on anticipated distribution shifts.

\raisebox{-0.3\height}{\hspace{0.05cm}\includegraphics[width=0.45cm]{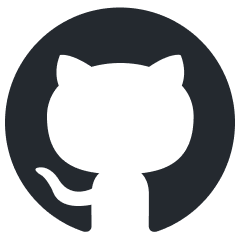}} \small \textbf{\mbox{Code:}} \href{https://github.com/Ashigarg123/ShiftySpeech/}{https://github.com/Ashigarg123/ShiftySpeech} \\
\vspace{1em}
\raisebox{-0.3\height}{\includegraphics[width=0.4cm]{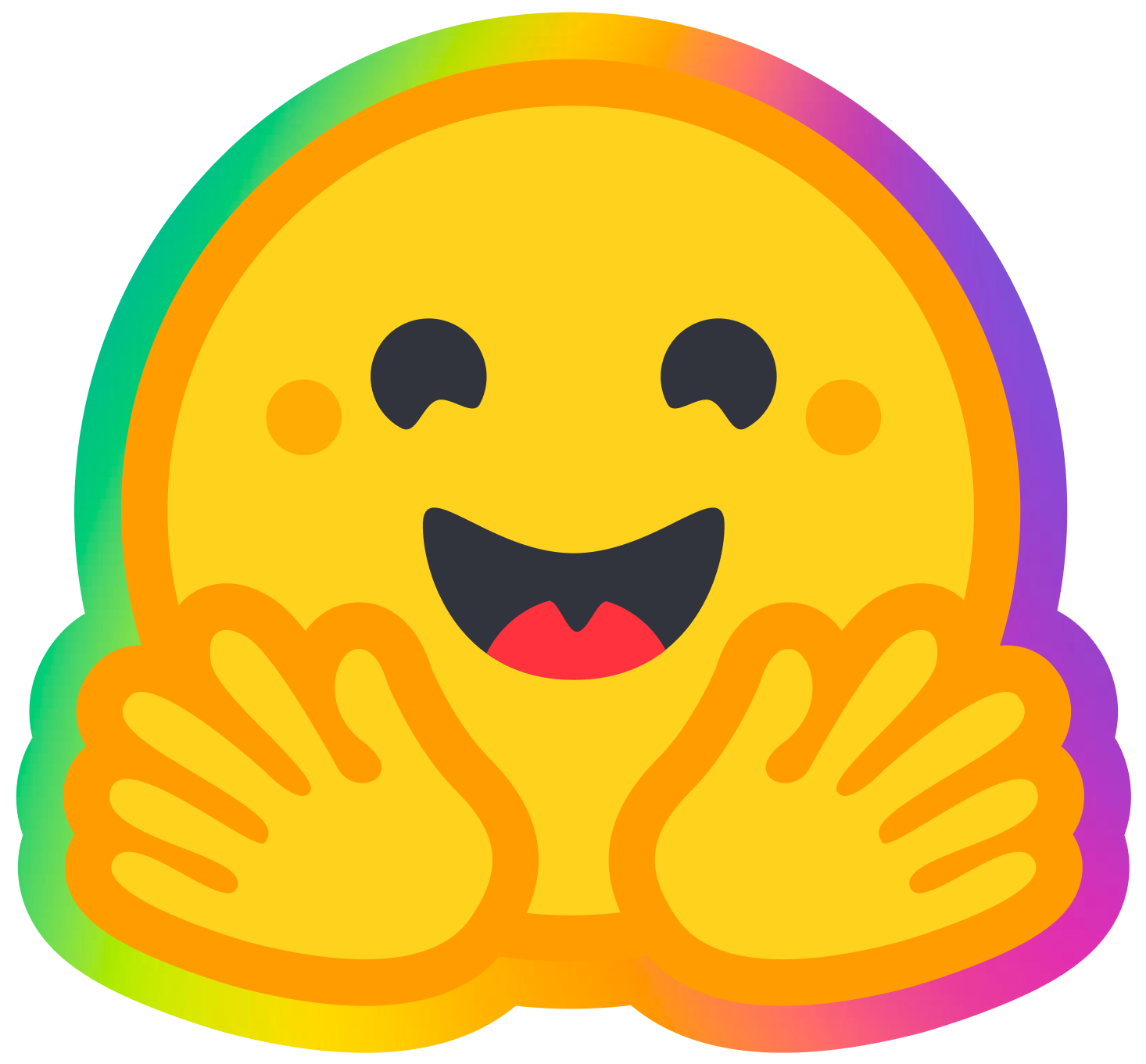}} \small \textbf{\mbox{Dataset:}} \href{https://huggingface.co/datasets/ash56/ShiftySpeech}{https://huggingface.co/datasets/ash56/ShiftySpeech}

\end{abstract}

\section{Introduction}\label{sec:intro}

\begin{figure}[htp]
 \caption{Illustration of distribution shift in ShiftySpeech. Train–test mismatches can arise from differences in language, speaker, or recording conditions.}
\vskip 0.2in 
\begin{center}
\centerline{\includegraphics[width=0.99\linewidth]{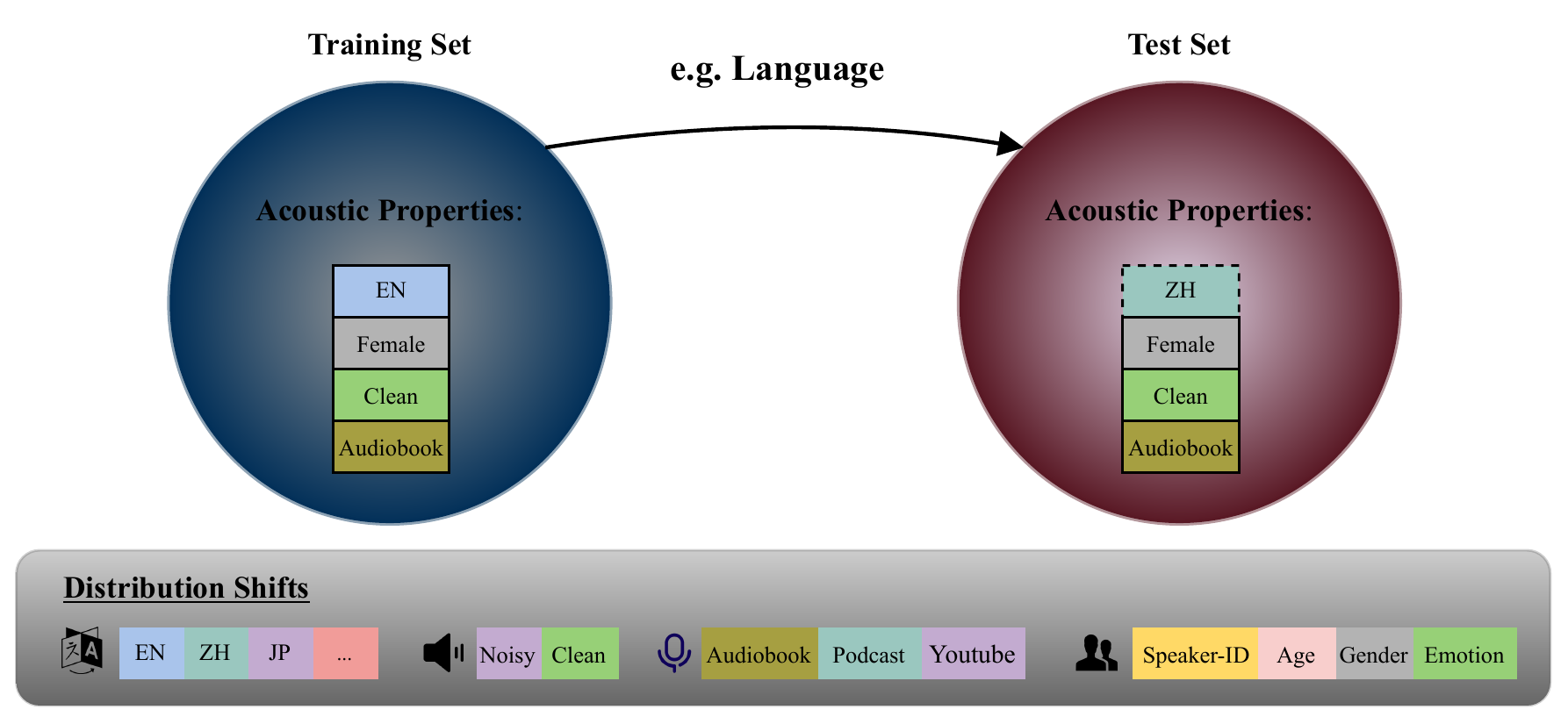}}
     \label{fig:dist-shifts}
\end{center}
\vskip -0.2in 
\end{figure}

Generalization of synthetic speech detectors is paramount ~\cite{muller2022does,chen20_odyssey}, which are increasingly deployed in security critical systems~\cite{kassis2023breaking} to detect synthetic speech from many speakers in the presence of varying acoustic conditions. Harder yet, synthetic speech detectors must generalize to novel spoofing attacks powered by the rapid pace of change in generative speech technology---including voice conversion (VC) and text-to-speech (TTS)---used in consumer products, such as e-readers, voice-assistants and chatbots~\cite{yang2024streamvcrealtimelowlatencyvoice, kameoka2018starganvcnonparallelmanytomanyvoice, choi2023dddmvcdecoupleddenoisingdiffusion, casanova2023yourttszeroshotmultispeakertts, qian2019autovczeroshotvoicestyle, kaneko2019cycleganvc2improvedcycleganbasednonparallel, bargum2023reimaginingspeechscopingreview}.

Numerous countermeasures have been increasingly proposed, reporting low Equal Error Rates (EERs) on benchmark datasets \cite{todisco2019asvspoof, yamagishi2021asvspoof,frank2021wavefakedatasetfacilitate}. These improvements are typically achieved either by expanding the training data to include more diverse domain conditions ~\cite{shim2023multidatasetcotrainingsharpnessawareoptimization} or to improve on ``hard'' data by increasing model capacity \cite{7993036,8398462,8812621,chen17_interspeech}.
Several work \cite{wang2022investigatingselfsupervisedendsspeech,kulkarni24_asvspoof} have also explored the effect of various front-end self-supervised learning (SSL) based models on the generalization of synthetic speech detectors using WavLM \cite{Chen_2022}, HuBERT \cite{hsu2021hubert} and Wav2Vec2 \cite{baevski2020wav2vec20frameworkselfsupervised}. 

However, evaluations in these works are typically limited to datasets with relatively narrow distribution shifts, thus limiting our systematic understanding of how well these systems generalize under distribution shifts that are likely to occur in real-world scenarios. To explore generalization in real-world settings, there is growing concern and corresponding need for newer datasets to assess robustness and generalizability \cite{muller2022does, müller2024harderdifferentunderstandinggeneralization,yi2023audiodeepfakedetectionsurvey,Cuccovillo_2022}. Although current benchmark datasets \cite{todisco2019asvspoof, yamagishi2021asvspoof,frank2021wavefakedatasetfacilitate} facilitate research in synthetic speech detection, they often lack diversity in domain conditions required to study generalization under real-world distribution shifts. While a limited number of datasets such as In-the-Wild \cite{muller2022does}, SpoofCeleb \cite{jung2025spoofcelebspeechdeepfakedetection} capture natural variability, they provide limited control over distribution shifts being evaluated to assess robustness (see \autoref{fig:dist-shifts}).

To enable more systematic evaluation, we develop a new benchmark database called \methodName, with over 3,000 hours of synthetic speech from 12 vocoders, 6 TTS systems, 3 languages, and 7 source domains. 
This benchmark database aims to explore the impact of a wide range of controlled distribution shifts on detector performance. Utilizing the proposed ShiftySpeech dataset, we aim to characterize the robustness of current state-of-the-art SOTA synthetic speech detectors much more generally and answer the questions: Are SOTA synthetic speech detectors robust to distribution shifts? Which distribution shifts, if any, degrade detection performance? 

\textbf{Main contributions:} (i) We create a benchmark dataset named ShiftySpeech comprising synthesized or vocoded speech from a variety of systems to help 
further research in robust detection under distribution shifts. (ii) We systematically study the generalization of SOTA synthetic speech detectors under the proposed distribution shifts. 
%(iii) We find that especially the complicated YouTube domain and unseen languages pose a significant challenge to our best performing detector. 

\section{Related Work}\label{sec:related_work}
\label{gen_inst}
\textbf{Distribution shifts} Machine learning models typically follow the statistical assumption that training and test data are independent and identically distributed (i.i.d.). However, this assumption often fails in real-world scenarios, where  significant shifts with respect to the training distribution are frequently encountered. This phenomenon, known as \textit{distribution shift}, has been widely studied in machine learning \cite{ geirhos2020generalisationhumansdeepneural} and can be primarily categorized into two major types---covariate shifts \cite{JMLR:v8:sugiyama07a,10.5555/2976456.2976474} and concept or conditional shifts \cite{10.1145/2523813}. The ShiftySpeech benchmark specifically addresses covariate shifts introduced by variations in TTS and vocoder systems, including shifts due to speaker characteristics, emotional expressiveness and language. While the standard covariate shift setup focuses on one source domain and one target domain, we focus on a more realistic scenario in which the test set may contain multiple unseen domains. 

\textbf{Re-vocoding} Training a generalizable detector considering various test-time shifts is often thought to require diverse training data generated using multiple TTS or VC systems, but this is computationally expensive and not always feasible at large scale. Among the datasets used for training synthetic speech detectors/countermeasures, data released as a part of ASVspoof challenge \cite{todisco2019asvspoof, yamagishi2021asvspoof} has been most commonly used. 
Recently released ASVspoof5 \cite{wang2024asvspoof5crowdsourcedspeech}, utilizes English subset of MLS \cite{Pratap_2020} as source, with increasing number of speakers. Recent work has also explored augmenting training data by generating additional synthetic samples through \textit{re-vocoding} (also known as copy-synthesis)---i.e., re-synthesizing real speech using vocoder-only pipelines that transform intermediate acoustic features (e.g., mel-spectrogram or F0) back into waveforms ~\cite{wang2023largescalevocodedspoofeddata, wang2023spoofedtrainingdataspeech, sun2023aisynthesizedvoicedetectionusing,sanchez14b_interspeech}. 

Apart from training, such re-vocoded data has been increasingly utilized for evaluating generalization performance \cite{zhang2024robustaisynthesizedspeechdetection,10769438}. While valuable, the aforementioned datasets remain limited in synthesis diversity---often relying on vocoders that have since become outdated as newer models emerged, and focusing primarily on English-language data. For instance, WaveFake \cite{frank2021wavefakedatasetfacilitate} majorly includes only GAN based vocoders and clean speech from LJSpeech \cite{ljspeech17} and JSUT \cite{sonobe2017jsut}, with limited speaker diversity, making it less suitable for test-time evaluations. LibriSeVoc \cite{sun2023aisynthesizedvoicedetectionusing} uses multi-speaker dataset (LibriTTS \cite{zen2019librittscorpusderivedlibrispeech}) as source and variety of vocoders, but is limited to English language and audiobook style recordings. In contrast, ShiftySpeech includes both re-vocoded and fully synthetic speech from recent models, spans multiple languages, and covers a diverse set of speakers and acoustic conditions. 

\textbf{Evaluating distribution shifts} While recent studies have reported low Equal Error Rates (EERs) on standard ASVspoof benchmark, several works have highlighted poor generalization in out-of-domain or real-world scenarios \cite{muller2022does,müller2024harderdifferentunderstandinggeneralization,ge2024spoofingattackaugmentationdifferentlytrained}. In order to address this, a range of new datasets have been proposed to better capture diverse real-world conditions: 
\begin{itemize}[itemsep=0mm, leftmargin=4mm, nosep]
    \item In-the-Wild (ITW) \cite{muller2022does} and Fake or Real (FoR) \cite{8906599} source the real data from a variety of public sources such as YouTube. ITW uses deepfakes of public figures found online, while FoR synthesizes speech using a range of TTS systems. However, both datasets are limited to English. 
    \item MLAAD \cite{muller2024mlaad} expands the scope to a multilingual setting, providing synthetic speech across 38 languages using various TTS systems.
    \item SpoofCeleb \cite{jung2025spoofcelebspeechdeepfakedetection} introduces greater speaker diversity by using VoxCeleb \cite{nagrani17_interspeech} as the source dataset and trains TTS models under real-world acoustic conditions rather than clean studio recordings to generate more realistic speech.
\end{itemize}
\begin{table}[!h]
\centering
\caption{Comparison of existing datasets and ShiftySpeech with respect to key characteristics for evaluating distribution shifts.}
\begin{adjustbox}{max width=\textwidth}
% \resizebox{\textwidth}{!}{%
\begin{tabular}{lcccccccccc}
\toprule
\textbf{Dataset} & 
\textbf{Size} & 
\makecell{\textbf{Source} \\ \textbf{Diversity?}} & 
\makecell{\textbf{TTS} \\ \textbf{Coverage?}} & 
\makecell{\textbf{Vocoder} \\ \textbf{Coverage?}} & 
\makecell{\textbf{Language} \\ \textbf{Coverage?}} & 
\makecell{\textbf{Speaker} \\ \textbf{Diversity?}} & 
\makecell{\textbf{Acoustic} \\ \textbf{Conditions?}} & 
\makecell{\textbf{Real-Synthetic} \\ \textbf{Pairs?}} & 
\makecell{\textbf{Controlled} \\ \textbf{Evaluation?}} \\
\midrule
% ASVspoof 2019  & -    & \xmark & \cmark & \xmark & \xmark & \cmark & \xmark & \xmark & \xmark \\
WaveFake       & 196h    & \cmark & \xmark & \cmark & \cmark & \xmark & \xmark & \cmark & \xmark \\
LibriSeVoc     &  118.08h    & \xmark & \xmark & \cmark & \xmark & \cmark & \xmark & \cmark & \xmark \\
MLAAD          &      420.7h      & \xmark & \cmark & \xmark & \cmark & \xmark & \xmark & \xmark & \xmark \\
In-the-Wild    & 17.2h          & \cmark & \xmark & \xmark & \xmark & \cmark & \cmark & \xmark & \xmark \\
Fake-or-Real   & Not available    & \cmark & \cmark & \xmark & \xmark & \cmark & \cmark & \cmark & \xmark \\
SpoofCeleb & Not available &\xmark & \cmark & \cmark &\xmark &\cmark & \cmark & \cmark & \xmark \\
\midrule
\textbf{ShiftySpeech} & \textbf{$>$3,000h} & \cmark & \cmark & \cmark & \cmark & \cmark & \cmark & \cmark & \cmark \\
\bottomrule
\end{tabular}
\end{adjustbox}
\label{tab:benchmark-comparison}
\end{table}
Despite these contributions, existing datasets often combine multiple source of variations, including recording conditions and speaker characteristics. Thus, making it difficult to isolate the impact of specific distribution shifts or synthesis methods. To address this limitation, we develop ShiftySpeech. It includes multiple controlled distribution shifts (e.g., language, emotion, background noise), provides parallel real and synthetic samples, and ensures consistent use of the same TTS and vocoder systems across all source datasets. This design enables systematic comparison and controlled experimentation across a range of realistic shifts.
\autoref{tab:benchmark-comparison} summarizes the existing datasets and ShiftySpeech
\section{ShiftySpeech: Benchmarking Distribution Shifts}\label{sec:data_generation}
% structure
\subsection{Overview}
The primary objective of ShiftySpeech is to provide the research community with a benchmark for evaluating the robustness of synthetic speech detection models under distribution shifts. The dataset is constructed through a systematic generation process that includes selecting source domains exhibiting natural variation in speaker characteristics, emotional expressiveness, language and acoustic conditions, using publicly available corpora \cite{wang2021voxpopulilargescalemultilingualspeech,chen2021gigaspeechevolvingmultidomainasr,sonobe2017jsut, bu2017aishell1opensourcemandarinspeech, ardila2020commonvoicemassivelymultilingualspeech,Lotfian_2019_3}(see \autoref{sec:dist_shifts_sources} for details). Synthetic speech is then generated using a broad range of modern TTS and vocoder systems (See \autoref{sec:data_gen}), enabling controlled experimentation across different types of generation mechanisms and distribution shifts. 
\begin{figure}[H]
 \caption{ShiftySpeech data generation pipeline illustrating three synthetic speech generation processes: text-to-speech (TTS), voice conversion (VC), and re-vocoding.}
\vskip 0.2in 
\begin{center}
\centerline{\includegraphics[width=0.99\linewidth]{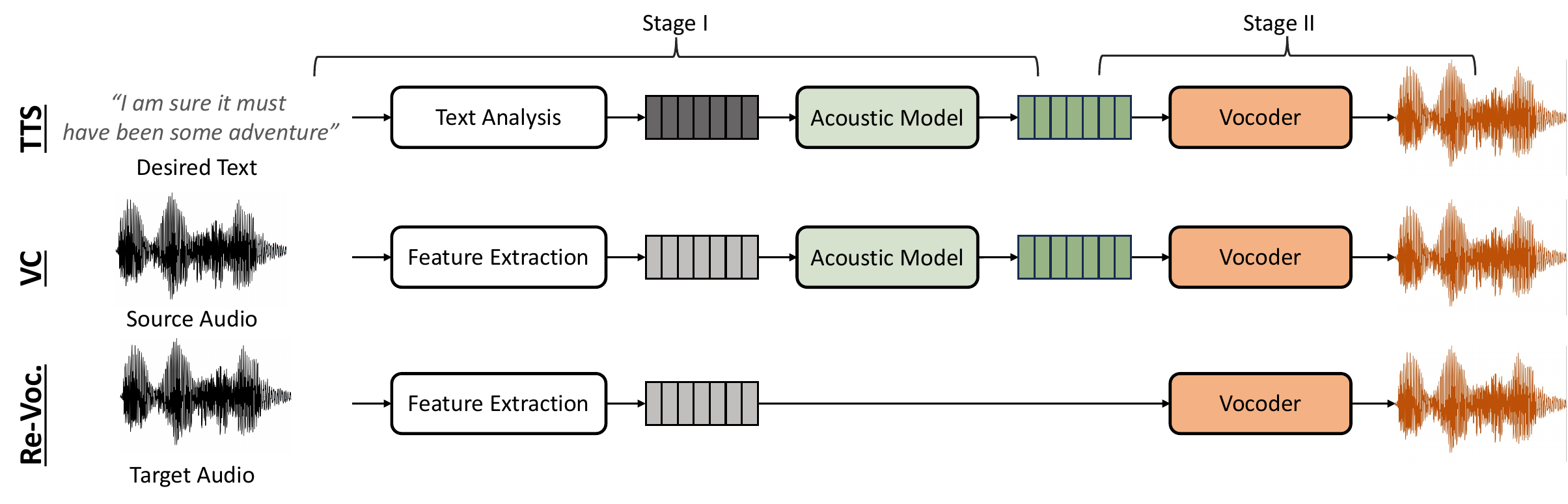}}
     \label{fig:shityspeech_pipeline}
\end{center}
\vskip -0.2in 
\end{figure}

\subsection{Source datasets in ShiftySpeech}\label{sec:dist_shifts_sources}
% % the above datasets are used to create distribution shifts 
Below, we describe source datasets used to introduce controlled distribution shifts for evaluating detection performance.

\textbf{Language:} For language shift, we use JSUT \cite{sonobe2017jsut} and AISHELL \cite{bu2017aishell1opensourcemandarinspeech} datasets. JSUT is a Japanese speech corpus, we resynthesize basic5000 subset of this dataset. It includes one female native Japanese speaker with audio in reading-style, comprising approximately 6.7 hours of audio. AISHELL is Mandarin speech corpus, with more than 170 hours of speech from 400 speakers.

\textbf{Celebrities / Speaking-style variation:} Here, the aim is to comprise the test set of varied speaking-style. We use VoxCeleb \cite{Chung18b} dataset for re-synthesis. It contains audios extracted from YouTube videos of various celebrities. In total 11 hours of audio samples are generated for each test vocoder.  

\textbf{Reading-style:} We utilize subset of LibriSpeech \cite{7178964} as the source data for reading-style audio. It consists of audio samples derived from publicly available audiobooks. 

\textbf{Emotion:} MSP-Podcast \cite{Lotfian_2019_3} is used as the source dataset. Audio samples in MSP-Podcast are collected using publicly available podcasts on various topics. 

\textbf{YouTube:} Audio samples are derived from YouTube subset of GigaSpeech \cite{chen2021gigaspeechevolvingmultidomainasr} dataset. This shift captures the presence of background noise and spontaneous speech. 

\textbf{Age, Gender and Accent:} Here, we re-synthesize subset of data from CommonVoice \cite{ardila2020commonvoicemassivelymultilingualspeech} such that it includes utterances from British and American accent. It includes seven age groups spanning from teens -- seventies. We cover four age-groups in our evaluations -- Twenties, Thirties, Forties and Fifties. Since, other subsets have relatively less number of samples they are not included in evaluations but we release the synthesized samples as part of the dataset. Please refer to \autoref{sec:exp-age-accent} for experimentation details.

\subsection{Data generation process}\label{sec:data_gen}
Figure~\ref{fig:shityspeech_pipeline} compares the generation processes of TTS, VC and re-vocoding. Speech synthesis is generally a two-stage process. Stage I deals with generating intermediate representations or acoustic features using source text or aligned linguistic features. These intermediate representations can be time-aligned features like mel-spectrograms and fundamental frequency (F0). Stage II deals with transforming these intermediate features to raw waveform or an audio. VC typically also follows a two-stage process, with the key difference being that it takes speech as input in Stage I, rather than text as in TTS.
The vocoder forms the core of Stage II in this pipeline and is widely used to generate synthetic audio, as it is less computationally expensive than fully end-to-end synthesis systems, as we mentioned in section \ref{sec:related_work}. ShiftySpeech consists of two subsets:

\vspace{-6pt}\paragraph{ShiftySpeech-TTS} For end-to-end TTS and VC systems, we include a variety of model types: VAE based VITS \cite{kim2021conditionalvariationalautoencoderadversarial}, non-autoregressive based FastPitch \cite{lancucki2021fastpitchparalleltexttospeechpitch}, flow-based Glow-TTS \cite{kim2020glowttsgenerativeflowtexttospeech}, diffusion-based Grad-TTS \cite{popov2021gradttsdiffusionprobabilisticmodel}, multi-speaker/multi-lingual methods (YourTTS \cite{casanova2023yourttszeroshotmultispeakertts} and XTTS \cite{casanova2024xttsmassivelymultilingualzeroshot}). 
Grad-TTS is sourced from their official website \footnote{\url{https://github.com/huawei-noah/Speech-Backbones/blob/main/Grad-TTS/README.md}}, while remaining models are implemented via the Coqui TTS framework \footnote{\url{https://github.com/coqui-ai/TTS/tree/dev}}. Based on those pretrained TTS/VC models, we synthsize data from diverse source data as shown in Table 
\ref{tab:shifty_speech_tts}. In detail, to support language diversity in synthesized speech, we utilize source transcripts from multiple reading-style speech datasets: LJSpeech \cite{ljspeech17} and VCTK \cite{yamagishi2019cstr} in English, AISHELL \cite{bu2017aishell1opensourcemandarinspeech} in Mandarin and JSUT \cite{sonobe2017jsut}in Japanese. 

\vspace{-6pt}\paragraph{ShiftySpeech-Voc} To generate vocoded data, we utilize publicly available pretrained checkpoints of vocoders trained on LibriTTS \cite{zen2019librittscorpusderivedlibrispeech} from their repositories as listed in Table \ref{tab:train_details_voc}. For vocoders without available checkpoints, we train vocoders based on LibriTTS from scratch. In the re-vocoding process, we firstly extract mel-spectrograms from real speech samples, then re-vocode it back to wavforms. We include seven source datasets to cover diverse distribution shifts as shown in Table \autoref{tab:shifty_speech_shifts}. 

UTMOS scores on ShiftySpeech dataset are reported in \autoref{tab:utmos-test-vocoders}. Other details on vocoder systems and training are provided in \autoref{sec:appendix-train-details}. 
\begin{table}[H]
\centering
    \caption{Distribution-shifts for TTS systems considered in this paper, along with the corresponding source datasets and duration. See \autoref{sec:tts-shift} for experiment details}
    \label{tab:shifty_speech_tts}
\vskip 0.15in
\begin{adjustbox}{max width=\textwidth}
  \begin{tabular}{lccc}
    \toprule
  \textbf{Distribution-shift} &  \textbf{TTS system} & \textbf{Source} & \textbf{Hrs}\\ 
    \midrule
    \multirow{3}{*}{TTS}  & VITS & \multirow{3}{*}{LJSpeech} &25.54h  \\
     &  Glow-TTS & & 25.70h  \\ 
      & Grad-TTS & & 22.64h \\
    \midrule
    TTS vs Vocoder &  LJSpeech trained HFG & LJSpeech & 24.36h \\
    \midrule
    Language & XTTS & JSUT & 7.54h \\ 
    \midrule
    \multirow{7}{*}{Language + TTS}    & VITS & \multirow{3}{*}{VCTK}&2.95h \\ 
    &  FastPitch &  &2.69h  \\ 
     & YourTTS &  & 3.47h  \\ 
     \cmidrule{2-4}
     &  \multirow{4}{*}{XTTS} & VCTK & 32.59h  \\
      & & AISHELL (dev) &19.99h  \\ 
     &  & AISHELL (test) &10.38h  \\
      & & AISHELL (train)&167.63h  \\
     \midrule
      \textbf{\textbf{Total hours}} & & & \textbf{345.48h}\\
         \bottomrule 
    \end{tabular}
\end{adjustbox}
\vskip -0.1in
\end{table}

\begin{table}[h]
\centering
    \caption{Distribution-shifts for vocoders considered in this paper. \texttt{ShiftySpeech} includes generated samples corresponding to each listed distribution-shifts for all test-time vocoders (\autoref{tab:train_test_vocoders}). Listed durations are per vocoder. For CommonVoice dataset following vocoders are excluded -- WaveGrad, APNet2 and iSTFTNet (see \autoref{sec:age_and_accent} for age, accent and gender related experiments).}
    \label{tab:shifty_speech_shifts}

    \vskip 0.15in
       \begin{adjustbox}{max width=\textwidth}
% \begin{sc}
    \begin{tabular}{lcr}
    \toprule
    \textbf{Distribution-shift}  & \textbf{Source} & \textbf{Hrs}\\ 
    \midrule
    Language (ja) & JSUT & 6.70h\\ 
    Language (zh) & AISHELL  & 10.00h\\ 
    Celeb / Speaking-style & VoxCeleb & 11.19h\\ 
    Reading-style & LibriSpeech  & 21.23h\\
    Emotive Speech; Spontaneous & MSP-Podcast  & 31.71h \\ 
    Background noise; Spontaneous; Near and far field & GigaSpeech-YouTube & 24.15h \\
    Age;Accent;Gender &CommonVoice  &246.00h \\
    \midrule
    \textbf{\textbf{Total hours (per vocoder)}} & & \textbf{350.98h} \\ 
      \bottomrule 
    \end{tabular}
\end{adjustbox}
\vskip -0.1in
\end{table}

\begin{comment}
\subsection{Quality assessment}
In order to assess the quality of speech samples generated, we report UTMOS \cite{saeki2022utmosutokyosarulabvoicemoschallenge} scores (See \autoref{sec:speech-quality}). UTMOS is a automatic mean opinion score (MOS) prediction system, proposed as a part of VoiceMOS challenge \cite{huang2022voicemoschallenge2022}. Mean Opinion Score (MOS) represents average human ratings of speech quality, ranging from scale of 1-5. \autoref{tab:utmos-test-vocoders} reports UTMOS score on ShiftySpeech dataset.
\end{comment}

In the following sections, we demonstrate the utilization of ShiftySpeech to study robustness under distribution shifts.

\section{Experiments}\label{sec:experiments}
Neural vocoders are being developed at a rapid pace and are widely used in conjunction with text-to-speech (TTS) systems, playing a key role in synthetic speech generation. Noting their critical role in the synthesis pipeline, vocoded speech serves as a promising direction for improved synthetic speech detection. This assumption is motivated by findings in prior work \cite{wang2023largescalevocodedspoofeddata}, suggesting vocoder artifacts help learn a good detection system, achieving low EER on ASVspoof and ITW benchmarks. Crucially, such artifacts are expected to persist even during test-time, enabling detectors trained on vocoded speech to generalize. \autoref{sec:exp-1} investigates this by evaluating detection performance across vocoder-induced distribution shifts in the ShiftySpeech benchmark. In \autoref{sec:tts-shift}, we extend this analysis to explore generalization when training includes TTS-generated data. \autoref{tab:train_test_vocoders} lists the vocoders used to synthesize training and evaluation data.

\begin{table}[h]
\centering
    \caption{Train and test time vocoders. Note that out of 16 vocoders, 3 are used both in training and as part of the evaluation set. Training details can be found in \autoref{sec:appendix-train-details}} 
    \label{tab:train_test_vocoders}
    \begin{adjustbox}{max width=\textwidth}
    \begin{tabular}{ll}
        \toprule
        \textbf{Split} & \textbf{Vocoders} \\
        \midrule
        Train-time & Mel, Mel-L, FB-Mel, WaveGlow \\
        \midrule
        \multirow{2}{*}{Test-time} & Style-Mel, WaveGrad, Vocos, BigVGAN, \\
        & BigVSAN, iSTFTNet, APNet2, UniVNET v1/v2 \\
        \midrule
        Train \& Test & PWG, HFG, MB-Mel \\
        \bottomrule
    \end{tabular}
    \end{adjustbox}
\end{table}

\subsection{Setups} 
\begin{comment}

\end{comment}

We train SSL-AASIST\footnote{\url{https://github.com/TakHemlata/SSL_Anti-spoofing}} \cite{tak2022automaticspeakerverificationspoofing} for the following experiments. It processes raw waveforms using wav2vec 2.0\footnote{\url{https://github.com/facebookresearch/fairseq/tree/main/examples/wav2vec}} XLSR-53 \cite{babu2021xlsrselfsupervisedcrosslingualspeech} \footnote{Although CommonVoice is one subset of the pretraining data for XLSR-53, we still include it in ShiftySpeech, as it is a large-scale dataset that introduces broader distribution shifts in age, accent, and gender.} as front-end and spectro-temporal graph attention network as backend \cite{jung2021aasistaudioantispoofingusing}. 
In order to evaluate performance on different systems, we use the standard Equal Error Rate (EER) metric. EER is defined as the value where the false acceptance rate equals the false rejection rate. Lower EER represents better detection performance.

\subsection{Distribution shifts and Impact of Vocoder Choice}\label{sec:exp-1}
To evaluate generalization under the distribution shifts included in ShiftySpeech-Voc, as introduced in the previous subsection, we use WaveFake as the training data and ShiftySpeech as the evaluation data. WaveFake \cite{frank2021wavefakedatasetfacilitate} is constructed by re-vocoding the LJSpeech dataset, which contains only a single speaker, making it a simpler and more controlled training source for comparison. We trained countermeasures in two ways as follows:

\vspace{-6pt}\paragraph{Trained on a single vocoder} We train seven separate detectors, each one trained on data generated from exactly one vocoder listed in the train-time column of Table \ref{tab:train_test_vocoders}. 
\autoref{tab:indv_voc} reports the average EER across 12 test-time vocoders for distribution-shifts present in ShiftySpeech. 

Detailed scores can be found in \autoref{sec:bes_vocoders}. We observe that for each of the distribution shifts, training on speech vocoded using the HiFiGAN (HFG) vocoder seems to be most effective, achieving the lowest average EER (aEER) of 9.14\%. However, training on WaveGlow alone leads to the highest aEER of 32.37\%. Results are summarized in \autoref{tab:indv_voc}.

\begin{table}[t]
\centering
    \caption{EERs (\%) on 12 test-time vocoders (see \autoref{tab:shifty_speech_shifts}) for each distribution-shift in ShiftySpeech. Models are trained on LJSpeech dataset with audio samples generated using each train vocoder.}
    \label{tab:indv_voc}
     \vskip 0.15in
    \begin{tabular}{lccccccc}
    \toprule
     & \multicolumn{6}{c}{\textbf{Test set}} \\
      \cmidrule(lr){2-8}
\textbf{Train set} & JSUT & AISHELL &VoxCeleb & Audiobook & Podcast & YouTube & aEER $\downarrow$\\ 
    \midrule
    WaveGlow &15.98 &27.36 &38.27 &34.22 &36.20& 42.23& 
   32.37\\
   Mel-L & 6.97 &20.83&19.81 &18.19&17.58&30.26 & 18.94\\
   Mel & 11.16 &21.36 &15.65 &16.86 &16.87& 26.09& 17.99\\
    PWG & 3.19 & 17.96& 18.12&18.38&21.58 &26.55 & 17.63\\
    FB-Mel &4.89  &12.61 &15.55 &16.18 &21.27& 27.14& 16.27\\
    MB-Mel & 3.35 &15.57 &12.98 &16.10 &17.14&22.46 & 14.60\\
   HFG & \textbf{1.22} &\textbf{10.67} &\textbf{8.40} &\textbf{9.08} &\textbf{8.35}&\textbf{17.17} & \textbf{9.14}\\
    \bottomrule 
    \end{tabular}
\vskip -0.1in
\end{table}

\vspace{-6pt}\paragraph{Leave-one-out} We further investigate if training on vocoded speech derived from multiple systems results in improved generalization. Detectors are trained by excluding one of the vocoders from the set of training vocoders. More details can be found in \autoref{sec:bes_vocoders}. Results from \autoref{tab:leave-one-out} indicate that holding out HFG and MelGAN-L degrades the performance on most of the distribution-shifts.  Further, based on the leave-one-out experiment, holding out HFG leads to the highest aEER of 14.79\%. While holding out MB-Mel, FB-Mel and WaveGlow improves generalization (aEER of 13.78\%). \autoref{tab:indv_voc} and \autoref{tab:leave-one-out} also highlight the relative difficulty of individual distribution-shifts. 

\vspace{-6pt}\paragraph{Discussion} Here, we note that SSL-based detectors are not uniformly robust to distribution shifts. In particular, certain shifts, such as the YouTube domain representing shift with respect to background noise and spontaneous speech, consistently degrades the detection performance. For instance, aEER on the JSUT subset of ShiftySpeech ranges from 1.22\% and 15.98\%, while for the Youtube subset it falls within a higher range of 17.17\% - 42.23\%. A similar trend was observed for the leave-one-out experiment with other distribution shifts, such as emotion (MSP-Podcast) and varying number of speakers (VoxCeleb) posing intermediate detection difficulty. 
Additional results on training the detector with a larger number of speakers (beyond the single-speaker setting) are presented in \autoref{sec:more_spks}. The relative impact of different distribution shifts remains consistent. For example, the YouTube domain continues to be the most challenging across both single and multi-speaker training setups.

\begin{table}[htb]
\centering
    \caption{Average EER (\%) on 12 test-time vocoders for each distribution-shift in ShiftySpeech. Models are trained with audio generated using multiple vocoders (excluding one), with LJSpeech as the source dataset.}
    \label{tab:leave-one-out}

 \begin{adjustbox}{max width=\textwidth}
    \begin{tabular}{lccccccc}
    \toprule
      & \multicolumn{6}{c}{\textbf{Test set}} \\
      \cmidrule(lr){2-8}
\textbf{Train set} & JSUT & AISHELL  &VoxCeleb & Audiobook & Podcast & YouTube & aEER $\downarrow$\\ 
    \midrule
       Leave HFG &2.18 & 19.73&12.46 &14.28 &14.96 &25.15 & 14.79\\ 
       Leave Mel-L &1.95 &21.08 &13.50 &12.42 &14.44 & 24.50 & 14.64\\ 
       Leave PWG &2.06 & 18.05&13.17&12.35 &14.54&26.75 & 14.48\\ 
       Leave Mel &2.07&16.82 &13.83 &12.49 &14.99&26.09 & 14.38\\ 
       Leave WaveGlow &1.95 &17.84 &12.13&11.47&13.39 &25.92 & 
        13.78\\ 
       Leave FB-Mel &2.12 &15.51 &12.78 &12.14 &13.75 & 25.22 & 13.58 \\ 
       Leave MB-Mel &1.93 &15.58 &11.89 &12.53 &13.90& 25.61 & 13.57\\ 
         \bottomrule 
    \end{tabular}
    \end{adjustbox}

\vskip -0.1in
\end{table}

\paragraph{Identifying challenging vocoders}\label{sec:chal_voc}
While we noted the generalization performance and specific distribution shifts in the previous section, here we further extend the previous experiments to identify vocoders particularly degrading overall generalization performance.
For each of the 7 detectors trained on synthetic speech generated using train-time vocoders in \autoref{tab:train_test_vocoders}, we note the average EER on each of the distribution-shifts and report the scores in \autoref{tab:harder_detect}. Scores on a particular detection system can be found in \autoref{sec:bes_vocoders}.

\begin{table}[H]
\centering
    \caption{For a given test vocoder, EER (\%) is averaged over scores from six train-time vocoders. See \autoref{tab:train_test_vocoders} for the list of train-time vocoders. See \autoref{sec:bes_vocoders} for a detailed score obtained using each train vocoder. Detection performance is worse on BigVGAN and best on PWG test vocoder. }
    \label{tab:harder_detect}
     \begin{adjustbox}{max width=\textwidth}

    \begin{tabular}{lcccccc|c}
    \toprule
    & \multicolumn{6}{c}{\textbf{Source}} & \\
\textbf{Test Vocoder}   & JSUT & AISHELL  &VoxCeleb & Audiobook & Podcast & YouTube & aEER $\downarrow$\\ 
    \midrule
    BigVGAN & 26.33 & 33.08 & 34.62 & 37.50 & 36.38 & 39.94 &34.64\\ 
    APNet2 & 10.01 & 25.08 & 24.79 & 26.82 & 26.56 & 33.01 &24.37 \\ 
    Vocos &  5.96 & 29.45 & 27.09 & 23.03 & 23.93 & 35.03 &24.08\\ 
    iSTFTNet & 10.11 & 20.43 & 20.46 & 22.52 & 23.23 & 29.79 & 21.09\\ 
    HFG & 10.20 & 19.03 & 20.83 & 22.07 & 24.20 & 26.68& 20.50 \\ 
    BigVSAN & 4.18 & 21.35 & 18.83 & 20.08 & 21.39 & 30.65 &19.41\\ 
    Univ-2 & 7.09 & 17.37 & 18.97 & 19.74 & 20.99 & 26.50 & 18.44\\ 
    Univ-1 & 3.88 & 8.79 & 15.67 & 15.63 & 17.23 & 23.93 &14.18\\ 
    WaveGrad  & 1.07 & 11.81 & 7.34 & 11.26 & 14.32 & 23.31 & 11.51\\ 
    Style-Mel & 1.02 & 11.09 & 11.14 & 11.06 & 13.15 & 21.38& 11.47 \\ 
    MB-Mel & 0.30 & 8.79 & 7.34 & 6.65 & 9.57 & 19.91 & 8.76\\ 
    PWG & \textbf{0.02} & \textbf{5.09} & \textbf{6.19} & \textbf{4.80} & \textbf{7.39} & \textbf{18.85} & \textbf{7.05}
   \\
   \bottomrule
    \end{tabular}
    \end{adjustbox}
  
\vskip -0.1in
\end{table}

\paragraph{Discussion} Detection performance on test vocoders varies significantly, with aEER values ranging from 7\% to 34\%. For example, PWG is relatively easier to detect, as all trained models achieve the lowest aEER on the PWG test set, with an average aEER of 7.05\%. On the other hand, BigVGAN is more challenging to detect, with higher EER values across all distribution-shifts and aEER of 34.64\%. Additionally, other vocoders show varying detection difficulty, resulting in intermediate aEER values. Overall, in addition to HFG, vocoders released in later years (iSTFTNet, APNet2, Vocos and BigVGAN) achieved high aEER in the range of 20\% -- 34\%. This depicts the generation quality being improved over the years. Appendix~\ref{sec:quality_and_detectability} provides further insight into whether more natural-sounding vocoders are harder to detect. 

\subsection{TTS distribution-shift}\label{sec:tts-shift}

In this section, we study distribution shifts introduced by fully end-to-end TTS systems from ShiftySpeech-TTS outlined in \autoref{tab:shifty_speech_tts}. Training a TTS system typically requires clean and studio-recorded speech. Therefore, generating data replicating real-world acoustic conditions with background noise, spontaneous speech and emotional expressiveness becomes challenging. 
Given this constraint, we mainly utilize ShitySpeech-TTS to discuss distribution shifts on language ~\cite{jung2025spoofcelebspeechdeepfakedetection}. 

Most detection systems are predominantly trained on English-language data. However, in real-world applications, these systems are often deployed in multilingual settings where test-time audio samples may span a wide range of languages. To systematically examine the impact of train-test language mismatch, we conduct two controlled experiments, discussed below.

\vspace{-6pt}\paragraph{Experimental Setup} For the following experiments, we use XTTS \footnote{\url{https://github.com/coqui-ai/TTS/tree/dev}} multi-lingual pre-trained model to generate training data. First, we train a detection model with generated utterances using the Mandarin dataset AISHELL as source. During test time we evaluate on English language multi-speaker dataset (VCTK). The same experiment is repeated by training on VCTK corpus and evaluating on AISHELL (zh) dataset. This setup allows us to study cross-language detection performance. 

\vspace{-6pt}\paragraph{Language + TTS shift} We utilize VCTK transcripts and Coqui TTS framework to generate the corresponding synthetic data. \autoref{tab:lang_tts_1} reports the generalization performance on various TTS systems, when a detector is trained on a SOTA TTS system (XTTS).

In \autoref{tab:lang_tts_1} we note that models trained with utterances generated from XTTS generalize well to utterances generated using VITS and YourTTS. However, these two models do not generalize comparatively as well to FastPitch.

\begin{table}[H]
\centering
    \caption{EER (\%) for models trained on XTTS-generated speech in Mandarin (AISHELL) and English (VCTK), evaluated under Language + TTS shifts and Language-only shifts.}
    \label{tab:lang_tts_1}
    \vskip 0.15in
   \begin{adjustbox}{max width=\textwidth}
        \begin{tabular}{l ccc | ccccccccc}
            \toprule
            \multirow{3}{*}{\textbf{Train Set}} & \multicolumn{3}{c|}{\textbf{Language + TTS Shift}} & \multicolumn{9}{c}{\textbf{Language-only Shift (XTTS)}} \\
            & \multicolumn{3}{c|}{VCTK (EN)} & VCTK & AISHELL & JSUT & \multicolumn{6}{c}{MLAAD} \\
            \cmidrule(lr){2-4} \cmidrule(r){5-5} \cmidrule(r){6-6} \cmidrule(r){7-7} \cmidrule(lr){8-13} 
            & YourTTS & VITS & FastPitch & EN & ZH & JA & PL & IT & FR & DE & ES & RU \\
            \midrule
            AISHELL-ZH & 0.42 & 1.86 & 13.73 & 0.18 & 0.00 & 0.00 & 0.20 & 2.90 & 4.80 & 5.40 & 4.90 & 11.50 \\
            VCTK-EN    & 0.20 & 0.00 & 0.44   & 0.02 & 0.04 & 0.06 & 1.20 & 6.70 & 6.60 & 5.40 & 7.90 & 10.40 \\
            \bottomrule
        \end{tabular}
        \end{adjustbox}
    \vskip -0.1in
\end{table}

\paragraph{Language-only shift} To further isolate the effect of language, we train and test using the same TTS system (XTTS), while varying only the language or source corpus at test time.

As shown in \autoref{tab:lang_tts_1}, models trained on Mandarin data (AISHELL) generalize well to English, and vice versa. We further extend the evaluation dataset to the Japanese language dataset, JSUT. Both models perform well on this dataset, with the model trained on AISHELL achieving perfect detection performance. 

We further expand the scope of this experiment by utilizing test data for more languages derived from MLAAD \cite{muller2024mlaad} dataset and corresponding real audio samples sourced from M-AILABS \cite{mai} dataset. \autoref{tab:lang_tts_1} reports the result on six different languages using detectors trained on English and Mandarin data. Surprisingly, the overall generalization of the model trained on Mandarin utterances is better than that of the model trained on English. While the performance trend is largely consistent across models, with languages such as Polish and German being easier to detect, Russian remains more challenging, showing an EER of 11.50\% for the AISHELL-trained model and 10.40\% for the VCTK-trained model.

\vspace{-6pt}\paragraph{Discussion} Although we were able to train the detection model using utterances synthesized from the Mandarin corpus, high-quality transcripts for many other languages are not readily available, limiting the ability to generate comparable training data. Our results suggest that training with non-English languages like Mandarin shows potential to improve generalization to unseen test-time languages. This indicates training on data beyond English is worth exploring to achieve better generalization performance.
To further complement our analysis, we also include a comparison between detectors trained on vocoder-based and TTS-based synthetic speech. Details are provided in Appendix~\ref{sec:tts_vs_vocoded}.

\section{Conclusion}\label{sec:conclusion}
\paragraph{Outlook} 
Although we observe significant degradations in detection performance under all tested conditions, certain shifts, such as the introduction of YouTube speech, have a more drastic impact on performance. 
While our findings highlight weaknesses of existing approaches to synthetic speech detection, we hope the proposed benchmark enable the development of more robust approaches.

For example, while our study focuses on the zero-shot setting, in certain applications it is reasonable to assume that a small sample of synthetic speech from the test conditions is available, which opens the door to building few-shot detectors that could adapt to evolving test conditions~\cite{soto2024fewshotdetectionmachinegeneratedtext}. Second, in settings where it is not feasible to perform few-shot adaptation, a possible recourse may be to consider anomaly detection approaches (e.g., \cite{ren2021simple}) that only seek to characterize in-distribution samples, making no assumptions about the out-of-distribution data.

\vspace{-6pt}\paragraph{Limitations} We perform systematic experiments with varying training and test conditions to understand the impact of distribution shifts on detection performance. However, due to space and computing limitations, we are naturally unable to consider all possible conditions. Specifically, we focus on a relatively narrow set of languages that are well-supported by existing synthesis methods, including English, Mandarin and Japanese. Finally, we choose to focus on a small number of state-of-the-art detector architectures that use self-supervised representations, specifically SSL-AASIST, as we expect this architecture to be the most robust to the kinds of distribution shifts we study here.

\small
\bibliographystyle{unsrtnat}
\bibliography{citations}

\newpage

\appendix

\section{Identifying best train-time vocoders}\label{sec:bes_vocoders}
The following tables provide detailed results corresponding to the experiments discussed in \autoref{sec:exp-1}. These tables record EER (\%) across detector systems and distribution-shifts.
\paragraph{Training details} Detection systems are trained on synthetic speech generated from a single vocoder using a learning rate of 1e-5 and batch size of 64 for 20 epochs. Detectors for leave-one-out experiments are trained for 50 epochs with a learning rate of 1e-6, batch size 64 and weight decay of 0.0001. Model checkpoint with lower validation loss is used for evaluation.

Average EER (aEER) column represents the overall generalization of each training model on given test vocoders. On the other hand, aEER row represents the overall difficulty of detecting a particular vocoder considering the performance of all detectors. 

\begin{table}[H]
\caption{EER (\%) on models trained on LJSpeech dataset and generated utterances from train-time vocoder. Evaluation data is generated using vocoded speech from test vocoders with JSUT (single-speaker) as the source dataset. Lowest aEER on all test vocoders is achieved using model trained with HFG vocoded samples. Higher aEER obtained with WaveGlow used in training.}
\label{tab:best_voc_jsut}
\vskip 0.15in
\begin{center}
\scriptsize
\resizebox{\textwidth}{!}{
\begin{tabular}{|c|c|c|c|c|c|c|c|c|c|c|c|c|c|}
\hline
Train set $\downarrow$ Test set $\rightarrow$ & \textbf{PWG} & \textbf{WaveGrad} & \textbf{BigVGAN} & \textbf{BigVSAN}& \textbf{MB-MelGAN}& \textbf{UniVNet v1}& \textbf{UniVNet v2} & \textbf{HiFiGAN} & \textbf{Style-MelGAN} & \textbf{Vocos} & \textbf{APNet2} & \textbf{iSTFTNet} &\textbf{aEER}\\
\hline
PWG & 0.0 & 0.24 & 20.40 &2.72 & 0.0  &0.26 &2.34  & 3.10& 0.0 &  2.28& 4.36& 2.64& 3.19\\
\hline
HiFiGAN & 0.0&0.10 & 10.48& 0.72&0.0 &0.0 &0.24 & 0.34& 0.06&0.70 &1.70 &0.36 &1.22
\\
\hline
MB-Mel &0.0 &0.32&20.50 &2.20 &0.0 &  0.50&  2.46& 2.62&0.02 &3.34 &5.02&3.26&3.35
\\
\hline
FB-Mel & 0.02& 0.56& 23.60 & 6.20 &0.0 & 1.00& 3.80& 4.36& 0.04&6.06 &8.36&4.68&4.89
\\
\hline
Mel & 0.02&3.54 & 34.58&7.60&0.88 & 4.40 & 11.10 &10.80 &0.58 & 14.00&28.78&17.70&11.16
\\
\hline
Mel-L & 0.0&  1.76&29.66 &4.64 &0.24 & 1.76& 5.28&6.84 &0.16 &  6.74&16.10&10.50&6.97
\\
\hline
WaveGLow& 0.12&0.98 &45.10 &5.20 & 0.98&19.24 & 24.44&43.36 & 6.28&8.64 &5.78&31.64&15.98
\\
\hline
\textbf{aEER} & 0.02& 1.07 & 26.33 & 4.18 & 0.30 & 3.88 & 7.09 &10.20 &1.02 &5.96 &10.01 & 10.11 & -\\ 
\hline
\end{tabular}}%
\end{center}
\vskip -0.1in
\end{table}

\begin{table}[H]
\caption{EER (\%) on models trained on LJSpeech dataset and generated utterances from multiple train-time vocoder in leave-one-out fashion. Evaluation data is generated using vocoded speech from test vocoders with JSUT (single-speaker) as the source dataset. Leaving Mel-l, MB-Mel and WaveGlow vocoder from training helps yield low aEER}
\label{tab:leave_voc_jsut}
\vskip 0.15in
\begin{center}
\scriptsize
\resizebox{\textwidth}{!}{
\begin{tabular}{|c|c|c|c|c|c|c|c|c|c|c|c|c|c|}
\hline
Train set $\downarrow$ Test set $\rightarrow$ & \textbf{PWG} & \textbf{WaveGrad} & \textbf{BigVGAN} & \textbf{BigVSAN}& \textbf{MB-MelGAN}& \textbf{UniVNet v1}& \textbf{UniVNet v2} & \textbf{HiFiGAN} & \textbf{Style-MelGAN} & \textbf{Vocos} & \textbf{APNet2} & \textbf{iSTFTNet} &\textbf{aEER}\\
\hline

Leave HFG &0.0  &0.08  &17.80& 1.06&0.0 &0.14 &0.92 & 1.30 &0.0& 1.20 & 2.08 &1.60&2.18
\\ 
\hline
Leave pwg & 0.0 &  0.16 &16.44&1.18 &0.0 &0.12 &0.82 & 1.14 &0.0&1.32  & 2.34&1.20&2.06
\\ 
\hline
Leave mel & 0.0 &0.16  &17.28&1.18 & 0.0&0.18 &0.78 & 1.08 &0.0&1.18  &  1.78&1.22&2.07
\\ 
\hline
Leave mel-l &0.0 &0.06  &16.06& 1.08 &0.0 &0.14 & 0.74&  1.16&0.02& 1.16 &  1.84&1.22&1.95
\\ 
\hline
Leave mb-mel & 0.0 &  0.12&15.66&1.12 &0.0 &  0.14& 0.78&  1.04&0.0& 1.14 & 2.22 & 1.04&1.93
\\ 
\hline
Leave fb-mel & 0.0 &0.18  &17.82&  1.12& 0.0&0.10 &0.82 &1.26  &0.0&1.16  & 1.80&1.22&2.12
\\ 
\hline
Leave waveglow & 0.0 &0.18  &15.58& 1.30 &0.0 &0.08 &0.68 &0.80  &0.02&  1.36& 2.54&0.92&1.95
\\ 
\hline
\end{tabular}}%
\end{center}
\vskip -0.1in
\end{table}

\begin{table}[H]
\caption{EER (\%) on models trained on LJSpeech dataset and generated utterances from train-time vocoder. Evaluation data is generated using vocoded speech from test vocoders with AISHELL (multi-speaker + accented) as the source dataset. BigVGAN and Vocos consistently exhibits higher EER across all training models.}
\label{tab:best_voc_aishell}
\vskip 0.15in
\begin{center}
\scriptsize
\resizebox{\textwidth}{!}{
\begin{tabular}{|c|c|c|c|c|c|c|c|c|c|c|c|c|c|}
\hline
Train set $\downarrow$ Test set $\rightarrow$& \textbf{PWG} & \textbf{WaveGrad} & \textbf{BigVGAN} & \textbf{BigVSAN}& \textbf{MB-MelGAN}& \textbf{UniVNet v1}& \textbf{UniVNet v2} & \textbf{HiFiGAN} & \textbf{Style-MelGAN} & \textbf{Vocos} & \textbf{APNet2} & \textbf{iSTFTNet}& \textbf{aEER}\\
\hline
PWG & 6.16&12.37  & 33.43 & 23.21& 9.69 & 10.53&  14.66 &20.07 &9.09 & 34.60 &  24.65&17.15&17.96\\
\hline
HiFiGAN & 2.24&8.41 &25.30 & 12.41 & 2.74& 3.05 &6.56 &8.47 &4.76 &31.36 & 16.04&6.80&10.67
\\
\hline
MB-Mel &4.66 &11.20&  34.37&19.34  & 5.07&8.94 &14.03 & 16.23&3.62 & 28.16&24.19&17.04&15.57
\\
\hline
FB-Mel &3.21 &6.55 & 31.96& 20.33 & 1.57&  5.32&  10.10& 10.99&3.13 &22.81 &22.32&13.08&12.61
\\
\hline
Mel &6.10 & 15.53&36.53 &24.63& 12.98&16.97  & 21.80 & 20.66& 11.87& 31.70&32.38&25.22&21.36
\\
\hline
Mel-L & 4.73 & 16.80& 37.38&26.93&12.57&11.87 &18.11 &22.68 &8.73&29.41 &34.37&26.47&20.83
\\
\hline

WaveGLow&8.57 &11.87&32.63 &22.24 &16.96 &42.12 & 36.39&  34.15&36.43 &28.17 &21.61&37.27&27.36
\\
\hline
\textbf{aEER} & 5.09 & 11.81 & 33.08 & 21.29 & 8.79 & 14.11 & 17.37 & 19.03 & 11.09 & 29.45 & 25.08 & 20.43 & - \\ 
\hline
\end{tabular}}

\end{center}
\vskip -0.1in
\end{table}

\begin{table}[H]
\caption{EER (\%) on models trained on LJSpeech dataset and generated utterances from multiple train-time vocoder in leave-one-out fashion. Evaluation data is generated using vocoded speech from test vocoders with AISHELL (multi-speaker + accented) as the source dataset. Leaving FB-Mel and MB-Mel vocoders from training helps yield low aEER }
\label{tab:leave_voc_aishell}
\vskip 0.15in
\begin{center}
\scriptsize
\resizebox{\textwidth}{!}{

\begin{tabular}{|c|c|c|c|c|c|c|c|c|c|c|c|c|c|}
\hline
Train set $\downarrow$ Test set $\rightarrow$& \textbf{PWG} & \textbf{WaveGrad} & \textbf{BigVGAN} & \textbf{BigVSAN}& \textbf{MB-MelGAN}& \textbf{UniVNet v1}& \textbf{UniVNet v2} & \textbf{HiFiGAN} & \textbf{Style-MelGAN} & \textbf{Vocos} & \textbf{APNet2} & \textbf{iSTFTNet}& \textbf{aEER}\\
\hline
Leave HFG & 9.93 &21.09  & 34.01&24.62 & 15.38&14.00 &17.89 &  19.98&3.66& 30.11&25.91&20.17&19.73
\\ 
\hline
Leave pwg &10.21  & 18.60 &33.20&22.43 & 13.54&10.95 &13.90 & 15.94&8.54&29.40 &24.70 &15.23&18.05
\\ 
\hline
Leave mel & 13.08 &16.17  &31.49& 20.05&12.83 & 9.88& 12.61& 15.49 &9.88& 26.28 & 20.56  &13.60
&16.82

\\ 
\hline
Leave mel-l & 15.56 & 23.83  & 32.84& 25.07& 18.07&17.64 &20.22 & 20.33 &4.23&30.64  & 24.34 &20.20&21.08
\\ 
\hline
Leave mb-mel & 7.24 &16.48  &30.08&19.14 &10.88 &8.82 &12.48 & 14.42 &4.43& 27.73 & 21.12 &14.15&15.58
\\ 
\hline
Leave fb-mel & 7.93 &15.81  &29.29& 18.96&11.09 &10.46 & 13.37& 14.32 &5.07& 27.06 & 19.89&13.51&15.51
\\ 
\hline
Leave waveglow &9.25  &19.98  &33.06&23.29 &13.26 & 10.82&  14.31& 16.06 &3.97& 30.72 & 24.05&15.30&17.84
\\ 
\hline
\end{tabular}}
\end{center}
\vskip -0.1in
\end{table}

\begin{table}[H]
\caption{EER (\%) on models trained on LJSpeech dataset and generated utterances from train-time vocoder. Evaluation data is generated using vocoded speech from test vocoders with VoxCeleb as the source dataset. Lowest aEER on all vocoders is achieved using model trained with HFG vocoded samples. Highest aEER obtained with WaveGlow used in training. }
\label{tab:bes_voc_voxceleb}
\vskip 0.15in
\begin{center}
\scriptsize
\resizebox{\textwidth}{!}{
\begin{tabular}{|c|c|c|c|c|c|c|c|c|c|c|c|c|c|}
\hline
 Train set $\downarrow$ Test set $\rightarrow$ & \textbf{PWG} & \textbf{WaveGrad} & \textbf{BigVGAN} & \textbf{BigVSAN}& \textbf{MB-MelGAN}& \textbf{UniVNet v1}& \textbf{UniVNet v2} & \textbf{HiFiGAN} & \textbf{Style-MelGAN} & \textbf{Vocos} & \textbf{APNet2} & \textbf{iSTFTNet} &\textbf{aEER}\\
\hline
PWG &  5.64&17.13  & 36.17 &20.76 & 5.88  & 12.66& 17.09 &20.90 & 9.35 &  29.31& 24.21&18.36&18.12\\
\hline
HiFiGAN &1.00 &  5.80& 23.49& 6.81& 1.04&4.22 &5.02 & 7.14&2.31 & 27.39&9.99 &6.70&8.40
\\
\hline
MB-Mel &2.15 &7.03&33.62 & 13.97& 1.95& 8.06 &12.12 &16.31 &  4.06&  20.84&19.94&15.67&12.98
\\
\hline
FB-Mel &4.06 &7.50 &37.64 & 19.75& 2.62& 10.15&15.59 & 17.19& 5.13& 24.10 & 23.82&19.10&15.55
\\
\hline
Mel &3.77 &  9.27&33.54 &12.72&6.13 & 11.26 & 16.31 & 18.79& 8.22&23.20 &27.37&17.23&15.65
\\
\hline
Mel-L &6.34 &15.08 &35.37 &20.61&10.13&14.42 &19.40 &23.71 &11.71& 24.74&33.52&22.75&19.81
\\
\hline
WaveGLow& 25.70&36.58& 42.51& 37.25&23.67 &48.97 &47.29 &41.81 &37.23 & 40.11&34.73&43.43&38.27
\\
\hline 
\textbf{aEER} & 6.95 & 14.05 & 34.62 & 18.83 & 7.34 & 15.67 & 18.97 & 20.83 & 11.14 & 27.09 & 24.79 & 20.46 & - \\ 
\hline
\end{tabular}}
\end{center}
\vskip -0.1in
\end{table}

\begin{table}[H]
\caption{EER (\%) on models trained on LJSpeech dataset and generated utterances from multiple train-time vocoder in leave-one-out fashion. Evaluation data is generated using vocoded speech from test vocoders with VoxCeleb as the source dataset. BigVGAN and Vocos consistently exhibits higher EER across all training models. }
\label{tab:leave_voc_voxceleb}
\vskip 0.15in
\begin{center}
\scriptsize
\resizebox{\textwidth}{!}{
\begin{tabular}{|c|c|c|c|c|c|c|c|c|c|c|c|c|c|}
\hline
 Train set $\downarrow$ Test set $\rightarrow$ & \textbf{PWG} & \textbf{WaveGrad} & \textbf{BigVGAN} & \textbf{BigVSAN}& \textbf{MB-MelGAN}& \textbf{UniVNet v1}& \textbf{UniVNet v2} & \textbf{HiFiGAN} & \textbf{Style-MelGAN} & \textbf{Vocos} & \textbf{APNet2} & \textbf{iSTFTNet} &\textbf{aEER}\\
\hline
Leave HFG & 3.48 &   8.71&31.33&13.66 &5.58 & 7.83&11.75 & 13.68 &4.00&19.45  &  16.72&13.39&12.46
\\ 
\hline
Leave pwg & 6.81 &   11.22&29.62&13.19 &8.24 & 8.08 &10.89 &  13.17 & 6.93& 21.50 & 15.65&12.74&13.17
\\ 
\hline
Leave mel &  8.30 &   10.99&31.04&14.77 & 8.45&9.06 &11.69 &13.62  &7.79&21.15  & 16.16 &12.96&13.83
\\ 
\hline
Leave mel-l &6.77  &  10.58&29.87&14.87 & 7.79 & 9.41&  11.85& 13.60 &6.48& 21.91 &  15.84&13.07&13.50
\\ 
\hline
Leave mb-mel & 4.45 & 9.39 &28.92& 11.96&6.15 & 6.85&9.76 &  12.18&4.82&  21.15& 15.34 &11.75&11.89
\\ 
\hline

Leave fb-mel & 6.19 &  11.12&28.66&13.09 & 7.53 & 8.39&11.16 & 12.66 &6.21& 21.44 &15.16 &11.75&12.78
\\ 
\hline
Leave waveglow & 6.07  &  9.47 &29.54&13.09 & 7.73&  6.66& 9.35& 11.85 &5.15& 20.39 &15.14 &11.22&12.13
\\ 
\hline

\end{tabular}}
\end{center}
\vskip -0.1in
\end{table}

\begin{table}[H]
\caption{EER (\%) on models trained on LJSpeech dataset and generated utterances from train-time vocoder. Evaluation data is generated using vocoded speech from test vocoders with Audiobook as the source dataset.  Lowest aEER on all test vocoders is achieved using model trained with HFG vocoded samples. Highest aEER obtained with WaveGlow used in training.}
\label{tab:best_voc_audiobook}
\vskip 0.15in
\begin{center}
\scriptsize
\resizebox{\textwidth}{!}{
\begin{tabular}{|c|c|c|c|c|c|c|c|c|c|c|c|c|c|}
\hline
Train set $\downarrow$ Test set $\rightarrow$ & \textbf{PWG} & \textbf{WaveGrad} & \textbf{BigVGAN} & \textbf{BigVSAN}& \textbf{MB-MelGAN}& \textbf{UniVNet v1}& \textbf{UniVNet v2} & \textbf{HiFiGAN} & \textbf{Style-MelGAN} & \textbf{Vocos} & \textbf{APNet2} & \textbf{iSTFTNet}&\textbf{aEER}\\
\hline
PWG & 4.37 & 14.53& 38.18 & 23.15& 4.96  & 13.91&19.85 & 22.06& 9.40 &23.02 & 25.55&21.67&18.38\\
\hline
HiFiGAN & 0.98 &  5.21& 26.12& 10.77 &0.75 &3.63 &6.30 &8.02 & 5.56 & 21.85 &12.05 &7.79&9.08
\\
\hline
MB-Mel & 2.77 &10.22& 37.63 &18.79 &3.40 & 11.88& 16.87& 19.90& 5.89 & 21.01&24.55&20.40&16.10
\\
\hline
FB-Mel & 2.89& 9.05& 38.46& 22.38&2.06 & 10.91&17.36 &18.91 &4.51 & 20.84&25.38&21.43&16.18
\\
\hline
Mel & 1.61& 7.84&38.75  &14.67 &5.31 &12.36  & 17.33 & 20.67& 7.88&20.76 & 33.93&21.22&16.86
\\
\hline
Mel-L & 2.73& 10.53& 38.45&19.67&7.40& 13.18&18.18 &22.46 & 8.70&20.79 &33.04&23.23&18.19
\\
\hline

WaveGLow& 18.31&21.47&44.94 & 31.14&22.71 &43.55 &42.34 & 42.50&  35.52& 32.95&33.30&41.91&34.22
\\
\hline
\textbf{aEER} & 6.23 & 11.26 & 37.50 & 20.08 & 6.65 & 15.63 & 19.74 & 22.07 & 11.06 & 23.03 & 26.82 & 22.52 & -\\
\hline
\end{tabular}}%
\end{center}
\vskip -0.1in
\end{table}

\begin{table}[H]
\caption{EER (\%) on models trained on LJSpeech dataset and generated utterances from multiple train-time vocoder in leave-one-out fashion. Evaluation data is generated using vocoded speech from test vocoders with Audiobook as the source dataset. Leaving HFG vocoder degrades the overall performance.}
\label{tab:leave_voc_audiobook}
\vskip 0.15in
\begin{center}
\scriptsize
\resizebox{\textwidth}{!}{

\begin{tabular}{|c|c|c|c|c|c|c|c|c|c|c|c|c|c|}
\hline
Train set $\downarrow$ Test set $\rightarrow$ & \textbf{PWG} & \textbf{WaveGrad} & \textbf{BigVGAN} & \textbf{BigVSAN}& \textbf{MB-MelGAN}& \textbf{UniVNet v1}& \textbf{UniVNet v2} & \textbf{HiFiGAN} & \textbf{Style-MelGAN} & \textbf{Vocos} & \textbf{APNet2} & \textbf{iSTFTNet}&\textbf{aEER}\\
\hline
Leave HFG &  2.88 &  7.46 &35.97&17.26 & 3.62&10.27 &15.37 & 16.94 &4.70& 17.36 &  21.18&18.40&14.28
\\ 
\hline
Leave pwg & 4.47 &  7.56&32.75& 14.75& 4.85&  7.45& 11.23&12.85 &5.53& 15.92 &17.16 &13.73&12.35
\\ 
\hline
Leave mel & 5.21  &  7.25 & 33.32&15.58 &5.41 & 7.45 & 11.82&  12.90 &5.41& 15.06 & 16.88 &13.69&12.49
\\ 
\hline
Leave mel-l & 4.10 &7.86 &33.13& 15.01& 4.59& 7.79& 11.94& 13.15 &3.56& 16.57 & 17.20 &14.14&12.42
\\ 
\hline
Leave mb-mel &3.26  &7.82  &33.92& 15.58& 3.66& 7.25& 11.89& 13.53 &4.10& 16.88 &  18.14&14.36&12.53
\\ 
\hline
Leave fb-mel & 3.88 &  7.57 & 33.24&14.42 &  4.40& 7.27& 11.43&12.66 & 5.46& 15.67 &16.62 &13.10&12.14
\\ 
\hline
Leave waveglow & 3.77 &  7.16 &32.31&14.72 &4.17 & 6.12& 9.94 &11.46  &3.61& 15.50 & 16.67&12.23&11.47
\\ 
\hline
\end{tabular}}
\end{center}
\vskip -0.1in
\end{table}
\hspace{0cm}

\begin{table}[H]
\caption{EER (\%) on models trained on LJSpeech dataset and generated utterances from train-time vocoder. Evaluation data is generated using vocoded speech from test vocoders with Podcast as the source dataset. Lowest aEER achieved using model trained using HFG vocoded samples. Higher aEER obtained with WaveGlow used in training.}
\label{tab:best_voc_podcast}
\vskip 0.15in
\begin{center}
\scriptsize
\resizebox{\textwidth}{!}{
\begin{tabular}{|c|c|c|c|c|c|c|c|c|c|c|c|c|c|}
\hline
Train set $\downarrow$ Test set $\rightarrow$ & \textbf{PWG} & \textbf{WaveGrad} & \textbf{BigVGAN} & \textbf{BigVSAN}& \textbf{MB-MelGAN}& \textbf{UniVNet v1}& \textbf{UniVNet v2} & \textbf{HiFiGAN} & \textbf{Style-MelGAN} & \textbf{Vocos} & \textbf{APNet2} & \textbf{iSTFTNet} &\textbf{aEER}\\
\hline
PWG &  9.34& 18.25 &38.70 & 25.64&  10.71 &17.20 &22.28  & 25.32& 14.45& 26.33 & 26.80&24.04&21.58\\
\hline
HiFiGAN &1.20 & 5.61& 24.46&9.53 & 1.34 &3.90 &5.78 & 8.81 &4.40 & 16.69& 11.34 &7.24&8.35
\\
\hline
MB-Mel & 4.57&11.55&36.55 &19.00 & 5.28& 13.13& 18.03& 22.71& 7.90&21.59 & 24.44&21.03&17.14
\\
\hline
FB-Mel & 7.23&15.63 &  40.16&27.29 & 7.04 & 16.88&23.06 & 25.23&10.19 &26.55 &29.96&26.10&21.27
\\
\hline
Mel & 3.25& 11.35& 34.72&15.64&  7.24&   12.65& 17.11 &21.31 & 8.99 & 20.96& 28.80&20.44&16.87
\\
\hline
Mel-L & 4.14& 12.73 &35.27 &18.85& 8.79& 11.50&16.43 & 22.40&9.23&20.41 &29.63&21.64&17.58
\\
\hline

WaveGLow&  22.00& 24.93& 44.82&  33.78&26.59 &45.39 & 44.29& 43.66&  36.91&35.00 & 34.99&42.13&36.20
\\
\hline
\textbf{aEER} & 7.39 & 14.29 & 36.38 & 21.39 & 9.57 & 17.23 & 20.99 & 24.20 & 13.15 & 23.97 & 26.56 & 23.23 & - \\ 
\hline
\end{tabular}}
\end{center}
\vskip -0.1in
\end{table}

\begin{table}[H]
\caption{EER (\%) on models trained on LJSpeech dataset and generated utterances from multiple train-time vocoder in leave-one-out fashion. Evaluation data is generated using vocoded speech from test vocoders with Podcast as the source dataset. Leaving WaveGlow vocoder from training helps yield low aEER}
\label{tab:leave_voc_podcast}
\vskip 0.15in
\begin{center}
\scriptsize
\resizebox{\textwidth}{!}{
\begin{tabular}{|c|c|c|c|c|c|c|c|c|c|c|c|c|c|}
\hline
Train set $\downarrow$ Test set $\rightarrow$ & \textbf{PWG} & \textbf{WaveGrad } & \textbf{BigVGAN} & \textbf{BigVSAN}& \textbf{MB-MelGAN}& \textbf{UniVNet v1}& \textbf{UniVNet v2} & \textbf{HiFiGAN} & \textbf{Style-MelGAN} & \textbf{Vocos} & \textbf{APNet2} & \textbf{iSTFTNet} &\textbf{aEER}\\
\hline
Leave HFG & 6.61 & 10.22 &31.94&16.93 & 7.37 & 11.57& 15.07& 17.76 &8.13& 17.78 & 18.95 &17.20&
14.96\\ 
\hline
Leave pwg &8.51  &   11.53&30.23&15.96 &9.17 &10.76 &13.50 &  16.02&9.58&17.36  & 17.06 &14.91 &14.54
\\ 
\hline
Leave mel & 9.85 & 11.75 &30.91&16.85 &10.12 & 11.16& 14.04& 15.98 &9.84&17.25  & 17.09 &15.05&
14.99\\ 
\hline
Leave mel-l & 8.14 &11.00  &29.84&16.34 &8.39 &11.09 &13.67 & 15.85 &8.82&18.06  &  16.85&15.29&14.44
\\ 
\hline
Leave mb-mel &7.20  & 10.67 &29.81&15.73 &7.77 &9.88 &13.03 &15.72  &8.16& 17.07 &17.13  &14.71&13.90
\\ 
\hline
Leave fb-mel &7.95  & 11.02 &29.02& 14.91&9.05 &10.64 & 12.81& 14.89 &8.98&16.28  &15.71 &13.83&
13.75\\ 
\hline
Leave waveglow & 7.36 & 10.46 &29.94& 15.51& 7.76 &  8.82& 11.76&14.29  &7.66& 16.94 &16.73 &13.44&13.39
\\ 
\hline
\end{tabular}}%
\end{center}
\vskip -0.1in
\end{table}

\begin{table}[H]
\caption{EER (\%) on models trained on LJSpeech dataset and generated utterances from train-time vocoder. Evaluation data is generated using vocoded speech from test vocoders with Youtube as the source dataset. Lowest aEER on all test vocoders is achieved using model trained with HFG vocoded samples. Highest aEER obtained with WaveGlow used in training.}
\label{tab:best_voc_yt}
\vskip 0.15in
\begin{center}
\scriptsize
\resizebox{\textwidth}{!}{
\begin{tabular}{|c|c|c|c|c|c|c|c|c|c|c|c|c|c|}
\hline
Train set $\downarrow$ Test set $\rightarrow$ & \textbf{PWG} & \textbf{WaveGrad V1} & \textbf{BigVGAN} & \textbf{BigVSAN}& \textbf{MB-MelGAN}& \textbf{UniVNet v1}& \textbf{UniVNet v2} & \textbf{HiFiGAN} & \textbf{Style-MelGAN} & \textbf{Vocos} & \textbf{APNet2} & 
\textbf{iSTFTNet}&
\textbf{aEER}\\
\hline
PWG &17.03 & 23.43 & 40.60 &32.45 &17.98 &20.75 &23.59 & 28.89& 18.24& 35.59 &31.71 &28.39&26.55\\
\hline
HiFiGAN & 9.45&13.29 &30.84&20.65&10.37 & 10.88& 11.51& 16.61& 10.33&37.11 & 19.17&15.92&17.17
\\
\hline
MB-Mel & 13.26&19.71& 39.86& 28.46& 14.97 & 18.54& 22.43&7.44 & 14.68&31.67 & 30.91& 27.59&
22.46\\
\hline
FB-Mel &16.17 & 23.83& 42.25&32.87 & 16.42& 22.87& 26.54&28.61 & 18.41&33.38 &33.94&30.39&27.14
\\
\hline
Mel & 17.66& 21.04&38.61&26.82&20.10 & 22.23 & 24.66&28.33 &19.16 & 31.59&34.86&28.03&26.09
\\
\hline
Mel-L &21.90 &26.80 & 41.45&33.01&25.33&25.40 &28.19 & 31.88&24.89& 33.83&38.47&32.02&30.26
\\
\hline
WaveGLow& 36.53&35.09& 45.99&40.30 & 34.25& 46.89& 48.58&45.02 &43.95 &42.05 & 42.02&46.20&42.23 \\
\hline 
\textbf{aEER} &18.85 & 23.31 & 39.94 & 30.65 & 19.91 & 23.93  & 26.50 & 26.68 & 21.38 & 35.03 & 33.01 & 29.79 & - \\ 
\hline
\end{tabular}}%
\end{center}
\vskip -0.1in
\end{table}

\begin{table}[H]
\caption{EER (\%) on models trained on LJSpeech dataset and generated utterances from multiple train-time vocoder in leave-one-out fashion. Evaluation data is generated using vocoded speech from test vocoders with Youtube as the source dataset. BigVGAN and Vocos consistently exhibits higher EER across all training models.}
\label{tab:leave_voc_yt}
\vskip 0.15in
\begin{center}
\scriptsize
\resizebox{\textwidth}{!}{
\begin{tabular}{|c|c|c|c|c|c|c|c|c|c|c|c|c|c|}
\hline
Train set $\downarrow$ Test set $\rightarrow$ & \textbf{PWG} & \textbf{WaveGrad} & \textbf{BigVGAN} & \textbf{BigVSAN}& \textbf{MB-MelGAN}& \textbf{UniVNet v1}& \textbf{UniVNet v2} & \textbf{HiFiGAN} & \textbf{Style-MelGAN} & \textbf{Vocos} & \textbf{APNet2} & 
\textbf{iSTFTNet}&
\textbf{aEER}\\
\hline
Leave HFG & 21.43 &22.96  &36.63& 26.73& 21.30 &  22.17&23.90 &25.10 &20.94& 27.99 &27.17&25.53&25.15
\\ 
\hline
Leave pwg &23.92  &25.62  &36.05&27.86 & 23.99 &24.06 & 24.86&  26.24&23.96& 30.24 & 27.98&26.26&26.75
\\ 
\hline
Leave mel & 22.85 &24.77  &36.59&27.98 &22.69 &23.17 &24.21 & 25.63 &22.80& 29.42 & 27.37 & 25.67&26.09
\\ 
\hline
Leave mel-l & 20.86 &23.36  &35.31&26.73 &20.36 &21.69 & 23.11 & 24.27 &19.06& 29.06 &  25.80&24.47&24.50
\\ 
\hline
Leave mb-mel & 22.58 & 24.24 &35.53& 26.74& 22.79&22.97 & 23.70 & 25.08 & 22.71&29.04  & 26.79 &25.20&25.61
\\ 
\hline
Leave fb-mel & 22.01 & 23.97 &35.00&26.63 & 21.76& 22.73& 23.77&24.71  &22.04& 29.03 & 26.34&24.73&25.22
\\ 
\hline
Leave waveglow & 23.18 & 24.90 &35.65& 27.06& 23.49&23.32 & 23.87&25.18  &23.03&  29.42& 26.91&25.13&25.92
\\ 
\hline

\end{tabular}}%
\end{center}
\vskip -0.1in
\end{table}
\section{Training details}\label{sec:appendix-train-details}
Below, we briefly describe the architectures of the GAN-based vocoders considered in this study

There are two distinct family of vocoders based on their nature of generation: Autoregressive (AR) \cite{oord2016wavenetgenerativemodelraw,mehri2017samplernnunconditionalendtoendneural,kalchbrenner2018efficientneuralaudiosynthesis} and Non-Autoregressive (NAR). Most of the new generation vocoder belongs to the latter family. Because AR generation includes sequential sampling and hence computationally inefficient with slow inference speed. Generative Adversarial Network (GAN) \cite{goodfellow2014generativeadversarialnetworks} based vocoders were then proposed to facilitate faster generation. GANs consists of a Generator and Discriminator network. A Generator maps noise vector to a data distribution (raw waveforms) and generate samples. While the Discriminator classifies a generated sample as real or fake with a probability. This includes training generator and discriminator simultaneously with generator minimizing the loss and discriminator maximizing it.
\paragraph{GAN-based vocoders:} We include variety of GAN-based vocoders. Archictetural details of these vocoders are discussed here in brief --- 
MelGAN (Mel) \cite{kumar2019melgangenerativeadversarialnetworks} follows GAN based architecture employing multiple window-based discriminators in order to learn better discriminative features from various frequency ranges. Parallel-WaveGAN (PWG) \cite{yamamoto2020parallelwaveganfastwaveform} uses 
NAR WaveNet based generator with joint adversarial and multi-resolution short-time Fourier Fourier transform loss (STFT). Multi-resolution STFT loss helps learn better time-frequency distribution for producing high quality waveform and also help stabilize GAN training and convergence. HiFiGAN (HFG) \cite{kong2020hifigangenerativeadversarialnetworks} further improved the generation quality of audio by including Multi-Period Discriminator (MPD) to model varied periodic patterns in an audio in addition to Multi-Scale Discriminator (MSD) like MelGAN. \cite{yang2020multibandmelganfasterwaveform} Multi-band MelGAN (MB-Mel) proposed an improved extension of MelGAN by expanding receptive field for improved quality. Instead of working with full-band signal, here generator network initially predicts sub-band level signals which are then combined to full-band signal and results in faster generation speed. \cite{mustafa2021stylemelganefficienthighfidelityadversarial} also uses sub-bands and employs temporal adaptive normalization (TADE) by facilitating conditional information about target speech to individual layers of general network. \cite{jang2021univnetneuralvocodermultiresolution} used MPD like HFG, in addition to multi-resolution spectrogram discriminator (MRSD) allowing for multiple spectrograms with different spectral and temporal resolutions. \cite{lee2023bigvganuniversalneuralvocoder} further improves both the generation quality and speed and used MRD. It further adds periodic inductive bias using Snake function. \cite{shibuya2024bigvsanenhancingganbasedneural} BigVSAN employs SAN \cite{takida2024saninducingmetrizabilitygan} instead of traditional GAN training. \cite{kaneko2022istftnetfastlightweightmelspectrogram,siuzdak2024vocosclosinggaptimedomain, ai2023apnetallframelevelneuralvocoder} iSTFTNet, VOCOS and APNet uses inverse short-time Fourier transform (iSTFT) in order to generate a waveform by predicting phase and amplitude. This helps avoid the need of several transposed convolutions to finally upsample input features to match the required sample rate. \cite{du2023apnet2highqualityhighefficiencyneural} APNET2 is an extension of APNET further adding improvements in phase prediction. 

In addition to above we also considered WaveGlow \cite{prenger2018waveglowflowbasedgenerativenetwork} as a Flow-based vocoder and  WaveGrad \cite{chen2020wavegradestimatinggradientswaveform} as a diffusion-based vocoder.

For all the vocoder systems listed below, training data used is LibriTTS. 

\begin{table}[H]
\caption{Training details of vocoders. For the pre-trained models the listed GitHub's are used with models trained on LibriTTS data.}
 \label{tab:train_details_voc}
 \vskip 0.15in
\begin{center}
\scriptsize
\begin{tabular}{lccc}
\hline
\toprule
Vocoder & GitHub  & Pre-trained \\
\hline
\toprule
PWG \multirow{4}{4em}& \multirow{3}{*}{\href{https://github.com/kan-bayashi/ParallelWaveGAN}{Link}} &  $\surd$  \\ 
MB-Mel & &  \\ 
Style-Mel & & \\ 
HiFiGAN & & \\ 
\hline
WaveGrad & \href{https://github.com/lmnt-com/wavegrad/tree/master}{Link}&  $\times$  \\ 
\hline 
UnivNET v1 \multirow{2}{4em}& \href{https://github.com/maum-ai/univnet}{Link}&  $\surd$ \\ 
UnivNET v2 &  & \\ 
\hline
BigVSAN & \href{https://github.com/sony/bigvsan}{Link}  & $\surd$ \\ 
\hline
iSTFTNet &\href{https://github.com/rishikksh20/iSTFTNet-pytorch/tree/master}{Link} & $\times$  \\ 
\hline 
APNet2 &\href{https://github.com/BakerBunker/FreeV/tree/main}{Link}  & $\times$  \\ 
\hline 
Vocos &\href{https://github.com/gemelo-ai/vocos/tree/main}{Link}  &  $\surd$ \\ 
\hline
BigVGAN & \href{https://github.com/NVIDIA/BigVGAN}{Link} & $\surd$  \\ 
\bottomrule
\end{tabular}
\end{center}
\vskip -0.1in
\end{table}

Following are the details for vocoders trained from scratch. WaveGrad is trained for 1M iterations with linear noise schedule, batch size of 64, learning rate of 2e-4 and sample rate of 24000. 
iSTFTNet is trained for 1.05M iterations with batch size of 240, learning rate of 0.0002, learning rate decay of 0.999, adam optimizer with $\beta_1$ as 0.8 and $\beta_2$ as 0.99
APNet2 is trained for 1M iterations with batch size of 480, learning rate of 0.0002, adam optimizer with $\beta_1$ as 0.8 and $\beta_2$ as 0.99
\begin{table}[H]
\caption{Training details of TTS systems used to generate the data.}
\label{tab:train_details_tts}
\vskip 0.15in
\begin{center}
\scriptsize
\begin{tabular}{lcc}
\hline
\toprule
TTS system & GitHub  & Pre-trained \\
\hline
XTTS &\multirow{3}{*}{\href{https://github.com/coqui-ai/TTS/tree/dev}{Link}} &  $\surd$  \\
VITS & & \\
YourTTS & &  \\ 
Glow-TTS & &  \\
\toprule
Grad-TTS & \href{https://github.com/huawei-noah/Speech-Backbones/blob/main/Grad-TTS/README.md}{Link}& $\surd$  \\ 
\bottomrule
\end{tabular}
\end{center}
\vskip -0.1in
\end{table}
A combination of multi-speaker and single-speaker models are used for TTS systems.

Detector trained on synthetic speech from TTS systems \autoref{sec:tts-shift} are trained with a learning rate of 1e-6, batch size of 128 and weight decay of 0.0001.

All experiments were conducted using an NVIDIA A100 GPU.

\section{New Generation vocoders}\label{sec:appendix-new-gen}

\paragraph{Training details} Models are trained for 15 epochs with a learning rate of 1e-6, batch size of 64 and weight decay of 0.0001. 
Test vocoders (see \autoref{tab:train_test_vocoders}) are divided based on their year of release. 
\begin{table}[H]
 \caption{Vocoders and year of release}
    \label{tab:voc_years}
    \vskip 0.15in
\begin{center}
\begin{small}
\begin{sc}
  \begin{tabular}{lc}
    \toprule
    \textbf{Vocoders} & \textbf{Year}\\ 
    \midrule
      WaveGlow & 2018\\
       \hline
       MelGAN and MelGAN-L & 2019\\ 
       \hline
       PWF, HFG and WaveGrad & 2020\\ 
       \hline
       MB-Mel, UniVNet and Style-Mel & 2021 \\
       \hline
       BigVGAN and iSTFTNet & 2022 \\
\hline 
APNet2 and Vocos & 2023 \\
\hline 
BigVSAN & 2024\\
         \bottomrule 
    \end{tabular}
    \end{sc}
\end{small}
\end{center}
\vskip -0.1in
\end{table}
\subsection{Exploring generalization on newly released vocoders}\label{sec:exp-3}

\autoref{sec:chal_voc} Notes for various distribution shifts, certain vocoders impact the generalization more as compared to others. Given the rapid pace of development, there is a significant risk that a detection system trained on older sets of generation methods is unable to distinguish speech synthesized using new methods from real speech. One possible solution to improve generalization in this scenario is to continuously update the training data. Can we improve detection performance by including new vocoders in the training data as they are released?

From our existing set of vocoders, we include vocoders released in 2018 in training and study the performance on vocoders released in the following years.\footnote{We use the publication year of the corresponding paper for these experiments, noting that this imperfectly captures the actual development time frame of the corresponding systems.} Similarly, we gradually include vocoders released in later years as part of training. Four detection systems are thus trained by including vocoders from 2018 -- 2021 sequentially. The test set includes out-of-domain vocoders released between 2022 to 2024. Detection performance on in-domain vocoders is additionally reported in \autoref{tab:detail_scores_gen_voc}. Information about vocoders and year of release can be found in appendix \autoref{tab:voc_years}.

The training data is derived from the WaveFake dataset. For vocoders not included in WaveFake, we generate fake audio samples using the respective vocoder with LJSpeech as the source. For evaluation, JSUT is used as the source dataset. Scores are reported in Table \ref{tab:gen_voc}. Detailed scores can be found in Table \ref{tab:detail_scores_gen_voc}. 

\begin{table}[ht]
    \caption{Models trained on samples generated using vocoders released in the year -- 2018 to 2021. Average EER is reported on samples generated using set of vocoders released in later years from 2022 to 2024 with JSUT dataset as source. Vocoders considered for each year can be found in \autoref{tab:voc_years}. aEER\% dropped significantly as new generation vocoders are added in training. The EER reduction is slow for vocoders released in year 2022. }
    \label{tab:gen_voc}
\vskip 0.15in
\begin{center}
\begin{small}
\begin{sc}
    \begin{tabular}{lcccccc}
    \toprule
Train set  & 2022 aEER & 2023 aEER & 2024 aEER \\ 
    \midrule
    2018 & 18.42 & 7.21 & 5.20 \\ 
    \hline
    2018 -- 2019 & 12.38 & 2.83 & 1.46 \\ 
    \hline
    2018 -- 2020 & 10.69 & 1.74 & 1.12 \\
    \hline
    2018 -- 2021 & 8.41 & 1.35 & 1.08 \\ 
    % \hline
    % HFG (best) & 5.42 & 1.20 & 0.72 \\ 
         \bottomrule 
    \end{tabular}
\end{sc}
\end{small}
\end{center}
\vskip -0.1in
\end{table}
Although there is continuous improvement in generalization performance, it can be noted that vocoders from 2022 pose a challenge to detection performance with low reduction aEER as compared to the 2023 evaluation set. In particular, this difficulty in generalization can be attributed to BigVGAN vocoder in evaluation set of 2022, which was previously found to be hard to detect (\autoref{sec:chal_voc}).
APNet2 and Vocos from 2023 evaluation set also achieved higher aEER on all distribution-shifts (see \autoref{tab:harder_detect}). However, this set achieved an overall better generalization performance with aEER of 1.35\%. This suggests that newer vocoders may still share similar artifacts with older ones, enabling the models to generalize effectively to them. Similar trends were observed for 2024 evaluation set; however, this test consists of only one vocoder and hence it might not be a good indicator of systems released in 2024. 

\begin{table}[H]
    \caption{EER (\%) reported for new generation vocoders. Detectors are inclemently trained by including vocoders released from successive year in the training set. Test vocoders released in year 2020 achieved low EER in general. Detection performance improves with more new generation vocoders in training. Model trained on HFG vocoded samples overall better generalization performance.}
    \label{tab:detail_scores_gen_voc}
    \vskip 0.15in
   
\begin{center}
 \begin{adjustbox}{max width=\textwidth}
    % \scriptsize
    % \resizebox{\textwidth}{!}{%
    \begin{tabular}{lcccccccccccc}
    \toprule
Train set $\downarrow$ Test set $\rightarrow$  & PWG 2020 & HFG 2020 & WaveGrad 2020 & MB-Mel 2021 & UniVNet c-16 2021 & UniVNet c-32 2021 & Style-Mel 2021 & BigVGAN 2022 & iSTFTNet 2022 & APNet2 2023 & Vocos 2023 & BigVSAN 2024  \\ 
    \midrule
    2018 & 0.12 & 43.36 & 0.98 & 0.98 & 19.24 & 24.44 & 6.28 & 5.20 & 31.64 & 5.78 & 8.64 & 5.20 \\ 
    \hline
    2018 -- 2019 & 0.14 & 4.08 & 0.24 & 0.16 & 0.82 & 2.30 & 0.22 & 21.10 & 3.66 & 3.38 & 2.28 & 1.46  \\ 
    \hline
    2018 -- 2020 & 0.12 & 1.54 & 0.08 & 0.12 & 0.24 & 1.06 & 0.08 & 19.90 & 1.48 & 2.08 & 1.40 & 1.12\\
    \hline
    2018 -- 2021 & 0.02 & 1.34 & 0.10 & 0.02 & 0.30 & 1.06 & 0.02 & 15.56 &  1.26 & 1.54 & 1.16 & 1.08  \\ 
    \hline
    HiFiGAN (best) &  0.0 & 0.34 & 0.10 & 0.0 & 0.0 & 0.24 &0.06 & 10.48 & 0.36 & 1.70 & 0.70 & 0.72  \\ 
         \bottomrule 
    \end{tabular} 
    % }%
    \end{adjustbox}
    \end{center}
    
\vskip -0.1in
\end{table}

\section{Training on Vocoded speech vs training on TTS speech}\label{sec:tts_vs_vocoded}
For this experimental setup, detectors are trained on LJSpeech dataset such that TTS system used to generate synthetic speech utilize HFG vocoder. For the TTS systems, we use the transcripts from LJSpeech dataset to generate corresponding samples. Each of the TTS system we utilize are pre-trained on LJSpeech dataset with HFG as the vocoder. Information about pre-trained models can be found in \autoref{tab:train_details_tts}. For the vocoder, we convert the genuine waveform into acoustic features and then reconstruct the waveform using the HFG vocoder trained on LJSpeech  \footnote{\url{https://huggingface.co/speechbrain/tts-hifigan-ljspeech}}. While we train individual detectors with different TTS systems, we also train detectors with synthetic speech consisting of just the vocoded speech. Additionally, we also train a system using both TTS + vocoded speech. Please note that the set of TTS systems used in training as well as evaluations are -- Grad-TTS, VITS and Glow-TTS. Evaluation data is a test split of LJSpeech dataset (in-domain). 

Detectors are trained with a learning rate of 1e-6, weight decay of 0.0001 and batch size of 64. In particular, the detectors trained on TTS speech are trained for 50 epochs, the detector trained on both TTS and vocoded speech is trained for 30 epochs, and the detector trained on vocoded speech is trained for 20 epochs.

\begin{table}[ht!]
    \caption{\textbf{TTS distribution-shift} Detectors are trained using generated samples from given TTS systems with transcripts from LJSpeech dataset. Poor generalization when the detection model is trained using TTS generated samples. Slight performance improvement when synthetic samples generated with HFG are used in training.}
    \label{tab:d_tts}
    \vskip 0.15in
\begin{center}

\scriptsize
\begin{sc}
    \begin{tabular}{lcccc}
    \toprule
    Training data & Grad-TTS & VITS & Glow-TTS & Vocoded \\ 
    \midrule
      Grad-TTS & 0.45& 38.55 & 1.90 & 38.32\\ 
        \hline
        VITS-TTS &43.81 &0.15  & 0.0 & 46.71 \\ 
        \hline
         Glow-TTS &44.58 & 31.60 & 0.0 & 45.34 \\ 
        \hline
      Vocoded-HFG  &28.39 & 24.50&0.0 & 0.0\\ 
      \hline
      TTS+Vocoded &0.30& 0.30 & 0.0 & 0.38 \\ 
         \bottomrule 
    \end{tabular}
      \end{sc}
\end{center}
\vskip -0.1in
\end{table}

Based on the results from \autoref{tab:d_tts}, it can be observed that training on synthetic speech generated using one end-to-end TTS system generalizes poorly to other end-to-end methods of generation as well as vocoded speech. For example, a model trained on synthetic audio samples from Grad-TTS achieved an EER of 38.55 \% on utterances generated from VITS dataset and 38.32 \% EER on vocoded speech from HFG. Similar can be observed for models trained on VITS and Glow-TTS. On the other hand training on just the vocoded speech led to a slight improvement in generalization with EER of 28.39 \% and 24.50 \% on Grad-TTS and VITS test datasets respectively. However, regardless of the generation system used to generate fake audio samples, Glow-TTS achieved an almost perfect detection score in all cases. 

We also compare the above results with one of the best models trained with fake audio samples generated from more than one vocoder as reported in \autoref{tab:leave-one-out}. EER dropped from 28.39 \% using just HFG vocoded speech to 23.43 \% using a detector trained with multi-source vocoded speech. Similarly, for VITS test data EER dropped to 20.45 \%. 
Training on both the TTS and Vocoded speech results in the best overall generalization results. This highlights the effect of distribution-shift with respect to different TTS systems during test time.

\paragraph{Discussion:} Poor generalization to unknown generation system reflects yet another test-time shifts to which detection models are vulnerable. This can also be regarded as one of most the common distribution-shift possible during test time. While the performance gain when the model is trained with fake and real samples generated using all the systems is significant, this is not an ideal real-world scenario. It can also be noted that performance gain using fake audio samples generated using vocoders is significant as compared to training using TTS generated samples.

\section{Age and accent}\label{sec:age_and_accent}
\subsection{Impact of age and accent}\label{sec:exp-age-accent}
In this experiment we investigate the effect of age, gender and accent as a distribution-shift. 

We evaluate the 7 trained models from \autoref{sec:exp-1} on CommonVoice dataset for 4 different age-groups (see \autoref{tab:age_groups_accent}). Note that evaluations are done on a subset of test vocoders belonging to new generation vocoder category (BigVGAN, BigVSAN, UniVNET v1, UniVNET v2 and Vocos).

\begin{table}[H]
  \caption{Detection performance for British accent speech and different age-groups. EER \% is averaged over seven train-time models (see \autoref{tab:age_groups_accent}). Test vocoders considered are BigVGAN, BigVSAN, UniVNET v1, UniVNET v2 and Vocos. Low aEER for test data with male speaker labels}
    \label{tab:bt_five_test_voc}
    \vskip 0.15in
\begin{center}
    \scriptsize
    \begin{tabular}{lccccc}
    \toprule
Gender $\downarrow$ Age $\rightarrow$  & Twenties & Thirties  & Forties & Fifties & \textbf{aEER}\\ 
    \midrule
Male & 22.02 & 21.67 & 23.87 & 23.41 & 22.74\\ 
\hline 
Female& 26.63 & 24.45 & 26.69 & 28.15 & 26.48\\
\bottomrule
    \end{tabular}
    \end{center}
\vskip -0.1in
\end{table}

\begin{table}[H]
 \caption{Detection performance for American accent speech and different age-groups. EER \% is averaged over seven train-time models (See \autoref{tab:age_groups_accent}). Test vocoders considered are -- BigVGAN, BigVSAN, UniVNET v1, UniVNET v2 and Vocos. Comparable aEER for test data with both male and female speaker labels}
    \label{tab:american_five_test_voc}
     \vskip 0.15in
\begin{center}
    \scriptsize

    \begin{tabular}{lccccc}
    \toprule
Gender $\downarrow$ Age $\rightarrow$  & Twenties & Thirties  & Forties & Fifties & \textbf{aEER}\\ 
    \midrule
Male & 22.09 & 21.79 & 24.98 & 26.16 & 23.75\\ 
\hline 
Female& 22.91 & 24.26 & 23.19 & 23.90 & 23.56\\
\bottomrule
    \end{tabular}
     \end{center}
\vskip -0.1in
\end{table}

We further extend the evaluation set of vocoders to include few more vocoders -- PWG, HFG, MB-Mel and Style-Mel. In these evaluations we utilize detector trained on HFG generated samples. Based on these experiments it can be observed that detection of audio samples with male speaker labels are comparatively easier to detect for both American and British accents. However this difference is more profound for British accented speech. For example, as per \autoref{tab:bt_hfg} there is approximately 29\% difference in performance for both genders. While approximately 11.4\% difference for American accented speech \autoref{tab:american_hfg}. Similar was observed based on \autoref{tab:bt_five_test_voc} and \autoref{tab:american_five_test_voc} with 14\% and 0.8\% difference between both genders with British and American accented speech. 
Moreover, age-group fifties were harder to detect in general for both the accents and genders considered.
It can be noted that performance on American accent is slightly better than British accent. We suspect this difference in performance stems from pre-training data used for wav2vec 2.0 xlsr. Moreover, even though the speaker utilized in training has female label, the detection performance was superior for evaluation data with male labels. This experiment again highlights the difference in generalization arising directly from pre-training data of employed front-end features. For better generalization, these differences should be taken into account while carefully selecting pre-training data.
\begin{table}[htbp]
 \caption{Detection performance for British accent speech and different age-groups. Test vocoders considered are -- BigVGAN, BigVSAN, UniVNET v1, UniVNET v2, Vocos, PWG, HFG, MB-Mel and Style-Mel. Detector trained with HFG vocoded samples is used for evaluation. Low aEER for test data with male speaker labels}
    \label{tab:bt_hfg}
    \vskip 0.15in
\begin{center}
    \scriptsize
    \begin{tabular}{lccccc}
    \toprule
Gender $\downarrow$ Age $\rightarrow$  & Twenties & Thirties  & Forties & Fifties & \textbf{aEER}\\ 
    \midrule
Male & 7.62 & 6.86 & 6.26 & 7.16 & 6.97\\ 
\hline 
Female&9.54 & 9.97 & 8.19 & 11.69 & 9.84 \\
\bottomrule
    \end{tabular}
    
    \end{center}
\vskip -0.1in
\end{table}

\begin{table}[ht!]
 \caption{Detection performance for American accent speech and different age-groups. Test vocoders considered are -- BigVGAN, BigVSAN, UniVNET v1, UniVNET v2, Vocos, PWG, HFG, MB-Mel and Style-Mel. Detector trained with HFG vocoded samples is used for evaluation. Low aEER for test data with male speaker labels}
    \label{tab:american_hfg}
    \vskip 0.15in
\begin{center}
    \scriptsize
    \begin{tabular}{lccccc}
    \toprule
Gender $\downarrow$ Age $\rightarrow$  & Twenties & Thirties  & Forties & Fifties & \textbf{aEER}\\ 
    \midrule
Male & 8.04 & 6.86 & 7.70 & 8.13 & 7.68\\ 
\hline 
Female& 8.32 & 9.09 & 8.02 & 9.27 & 8.67\\
\bottomrule
    \end{tabular}
    \end{center}
\vskip -0.1in
\end{table}
 
\begin{table}[H]
\caption{Scores for Each System Across Different Age Groups, Accents, and Genders. Scores are averaged over five test vocoders for simplicity}
\label{tab:age_groups_accent}
\vskip 0.15in
\begin{center}
\scriptsize
\begin{tabular}{|c|c|c|c|c|c|c|c|c|c|}
\hline
\multirow{3}{*}{\textbf{Train Vocoder}} & \multirow{3}{*}{\textbf{Accent}} & \multirow{3}{*}{\textbf{Gender}} & \multicolumn{4}{c|}{\textbf{Age Groups}} \\ \cline{3-7} 
 &  &  &  \textbf{Twenties} & \textbf{Thirties} & \textbf{Forties} & \textbf{Fifties} \\ \hline
\multirow{4}{*}{\textbf{PWG}} & \multirow{2}{*}{British} & Female & 28.69  &25.58  & 26.18 & 32.37 \\ \cline{3-7} 
 &  & Male &  21.51& 24.87 & 23.71 & 22.21 \\ \cline{3-7} 
 & \multirow{2}{*}{American} & Female &26.13  & 26.31 &26.47  & 27.30  \\ \cline{3-7} 
 &  & Male &  24.04 &22.03 &25.08  &29.03\\ \cline{3-7} 
 \hline
\multirow{4}{*}{\textbf{HiFiGAN}} & \multirow{2}{*}{British} & Female  & 13.52 & 14.11 &  11.52&15.86 \\ \cline{3-7} 
 &  & Male & 11.15 &  9.96& 9.41 &  10.42\\ \cline{3-7} 
 & \multirow{2}{*}{American} & Female &12.09  & 12.62 &11.38  &12.98 \\ \cline{3-7} 
 &  & Male  &11.09  &9.94  &10.23 &11.38 \\ \cline{3-7}  \hline
\multirow{4}{*}{\textbf{MB-Mel}} & \multirow{2}{*}{British} & Female & 24.38 & 20.78 & 24.42 & 20.17 \\ \cline{3-7} 
 &  & Male & 17.72 & 18.06 &  18.82&  19.64\\ \cline{3-7} 
 & \multirow{2}{*}{American} & Female&22.41 &22.52  &18.54  & 20.97\\ \cline{3-7} 
 &  & Male &16.34  &16.91  & 20.45 &22.44 \\ \cline{3-7} \hline
\multirow{4}{*}{\textbf{FB-Mel}} & \multirow{2}{*}{British} & Female & 27.53 & 25.70 &31.45 &27.72 \\ \cline{3-7} 
 &  & Male & 18.23 & 17.15 &22.50  &  22.71\\ \cline{3-7} 
 & \multirow{2}{*}{American} & Female  &22.23  &21.62  & 21.59 &25.19 \\ \cline{3-7} 
 &  & Male & 20.13 &19.71  & 24.40 &26.02  \\ \cline{3-7}  \hline
\multirow{4}{*}{\textbf{Mel}} & \multirow{2}{*}{British} & Female & 21.92 &17.28  & 21.62 & 25.47   \\ \cline{3-7} 
 &  & Male & 19.46 & 19.75 & 20.61 & 20.22  \\ \cline{3-7} 
 & \multirow{2}{*}{American} & Female &  19.12&19.60  &19.19  &18.74 \\ \cline{3-7} 
 &  & Male  &18.34  &18.52  &21.00  &21.60\\ \cline{3-7}  \hline
\multirow{4}{*}{\textbf{Mel-L}} & \multirow{2}{*}{British} & Female  & 21.59 &18.19  & 21.59 &26.53  \\ \cline{3-7} 
 &  & Male &19.36  &18.50  & 23.29 &  21.06\\ \cline{3-7} 
 & \multirow{2}{*}{American} & Female  &18.66  &21.72  &19.34  &18.68  \\ \cline{3-7} 
 &  & Male &19.58  &19.61  &24.89 & 23.89 \\ \cline{3-7}  \hline
\multirow{4}{*}{\textbf{WaveGlow}} & \multirow{2}{*}{British} & Female &48.79  & 49.55 & 50.07 & 48.99 \\ \cline{3-7} 
 &  & Male & 46.72 &  43.45& 48.80 &  47.63 \\ \cline{3-7} 
 & \multirow{2}{*}{American} & Female  &39.74  &45.47  &45.84  &43.45  \\ \cline{3-7} 
 &  & Male & 45.16 &45.77  &48.83  &48.80   \\ \cline{3-7}  \hline
\end{tabular}
\end{center}
\vskip -0.1in
\end{table}
For additional test purposes, we also include synthetic speech generated using end-to-end TTS systems using LJSpeech and VCTK transcripts.
 
\section{Speech Quality and Detectability}\label{sec:quality_and_detectability}

\subsection{Relation between speech quality and vocoder detectability}\label{sec:speech-quality}
In this section, we conduct experiments to explore the correlation between quality score (as reported by automated metric UTMOS \cite{saeki2022utmosutokyosarulabvoicemoschallenge}) and detection performance (as measured by EER).
For each of the datasets, we get UTMOS score on all 12 test vocoders. In addition we get the EER for test vocoders by averaging the EER performance obtained on the 7 train vocoders. We particularly utilize stronger learner model proposed as a part of UTMOS metric. This strong learner uses wav2vec2.0 pretrained on LibriSpeech as SSL model. Finally, we calculate the co-relation coefficient using Pearson Correlation. Scores are reported in \autoref{tab:utmos_eers}. We note a strong positive correlation between between two metrics for Audiobook and Podcast datasets with correlation score as 0.89 and 0.85 respectively. While there is moderate correlation between automatic quality metric and EER for VoxCeleb dataset with correlation score of 0.74. In addition, Youtube dataset also exhibits a positive relationship, but not very strongly with correlation score of 0.58. Further, we note that for hard to detect vocoders -- \{BigVGAN, HFG and Vocos\} UTMOS score is in higher range with higher EER (See \autoref{tab:utmos-test-vocoders}). While for compartively easier to detect vocoders like PWG, UTMOS scores are in moderate range of 2 -- 3.

\begin{table}[ht!]
    \small
    \caption{Co-relation score between UTMOS and EER, calculated using Pearson Correlation. All of the 12 test vocoders were considered to get UTMOS score on respective dataset. EER is averaged over score from 7 trained models. Moderate co-relation observed for YouTube and VoxCeleb}
    \label{tab:utmos_eers}
  \vskip 0.15in
\begin{center}
    \begin{tabular}{lcc}
    \toprule
Dataset & Score \\ 
\toprule
Audiobook & 0.89 \\ 
\hline
Podcast & 0.85 \\ 
\hline
YouTube & 0.58 \\ 
\hline
VoxCeleb & 0.74 \\ 
 \bottomrule 
    \end{tabular}
    \end{center}
\vskip -0.1in
\end{table}
\begin{table*}[h!]
 \caption{UTMOS scores on each dataset. Please note: UTMOS score calculation requires sampling rate to be 16kHz.}
    \label{tab:utmos-test-vocoders}
\vskip 0.15in
\begin{center}
    \scriptsize
    \begin{tabular}{clccccccc}
    \toprule
    \multicolumn{2}{c}{\multirow{2}{*}{\textbf{Vocoder}}} & \multicolumn{5}{c}{\textbf{UTMOS} $\uparrow$} \\
    \cmidrule(lr){3-9}
    & & Audiobook & MSP-Podcast & Youtube & Voxceleb & JSUT & AISHELL&CommonVoice \\
    \midrule
PWG && 3.00   &  2.32&2.01 &2.17 & 2.75& 1.72&  2.71\\ \hline
HiFiGAN && 3.56  &2.72  & 2.24&2.47 &3.44 & 1.76& 3.11\\ \hline
MB-MelGAN& & 2.99  & 2.29 &2.01 &2.03 & 2.72 &1.57 & 2.68\\ \hline
Style-MelGAN&&3.25   &2.51  &2.15 & 2.34 & 2.89 & 1.82& 2.89 \\ \hline
BigVGAN&&3.67   & 2.78 &  2.27& 2.60 &3.62  &1.74 &3.21\\ \hline
BigVSAN&&3.43   & 2.58 & 2.16& 2.39&3.35  &1.61 &3.03\\ \hline
UniVNet v1 &&  3.49  & 2.69  & 2.25&2.50 &3.35  &1.96 &3.13 \\ \hline
UniVNet v2 & & 3.57   & 2.76 & 2.29& 2.55&3.50  &1.96 &3.17 \\ \hline
WaveGrad &&3.20  & 2.42 &2.08 & 2.20&3.11  &1.54 &--\\ \hline
APNet2 && 3.56 &2.68& 2.22 &  2.48&3.45  &1.70 &-- \\
\hline
Vocos &&3.46 &2.59  &2.14 &  2.29&3.36&1.63 &2.91\\
\hline
iSTFTNet &&3.50& 2.65 & 2.22&  2.45 &3.43 & 1.81&--\\
         \bottomrule
    \end{tabular}
    \end{center}
\vskip -0.1in
\end{table*}
\paragraph{Discussion} The results suggest that automatic speech metrics can serve as a proxy for diagnosing fake audio samples, given their strong correlation with EER across most distribution sets. However, these metrics have limitations, particularly in providing consistent results under different conditions. While we adopt UTMOS as the objective metric for datasets in our experiments, exploring other automatic MOS prediction methods could provide valuable insights in the future. For instance, UTMOS doesn't seem to generalize well to AISHELL. We confirm this by getting a relatively low mean UTMOS score of 1.99 on real or bonafide version of AISHELL.

\section{Impact of training on more speakers} \label{sec:more_spks} 
In general, exposing a detection system to a variety of data during training is expected to help generalization. For example, including speakers with various speaking styles, pitch, timbre, and so on, may help generalize well to unseen speakers. Even though it is possible to access multiple-speakers during training, the number of audio samples per speaker are usually limited. In the following experiments, we explore the extent to which increasing the number of speakers in training helps generalization in this setting. 

\paragraph{Setup} In order to understand the importance of including different numbers of speakers in training, we increase the number of speakers from 1 to 10 in training while keeping the total number of training samples the same. For each setting, we repeat the experiment 5 times selecting different sets of speakers at random to account for possible variance due to speaker properties. For instance, when training a model with 5 speakers, we randomly select 5 speakers for each of the 5 experiments and train 5 different detection systems. Each of the 5 experiments builds on the previous one, with the 5-speaker training set expanding on the 4-speaker set by adding one new speaker. This overlap ensures that performance gains are solely due to the newly added speakers. 

Models are trained on the \texttt{train-clean-360} subset of LibriTTS and synthetic speech is generated using HFG vocoder. The detection models were trained for 40 epochs with a learning rate of 1e-5 and batch size 64, selecting the model with the lowest validation loss. 

\subsection{Performance on in-domain vocoder}
Detectors are trained on audio samples from HFG vocoder and evaluated on samples from the same vocoder.

\begin{figure}[H]
\vskip 0.2in 
\begin{center}
\centerline{\includegraphics[width=0.8\textwidth]{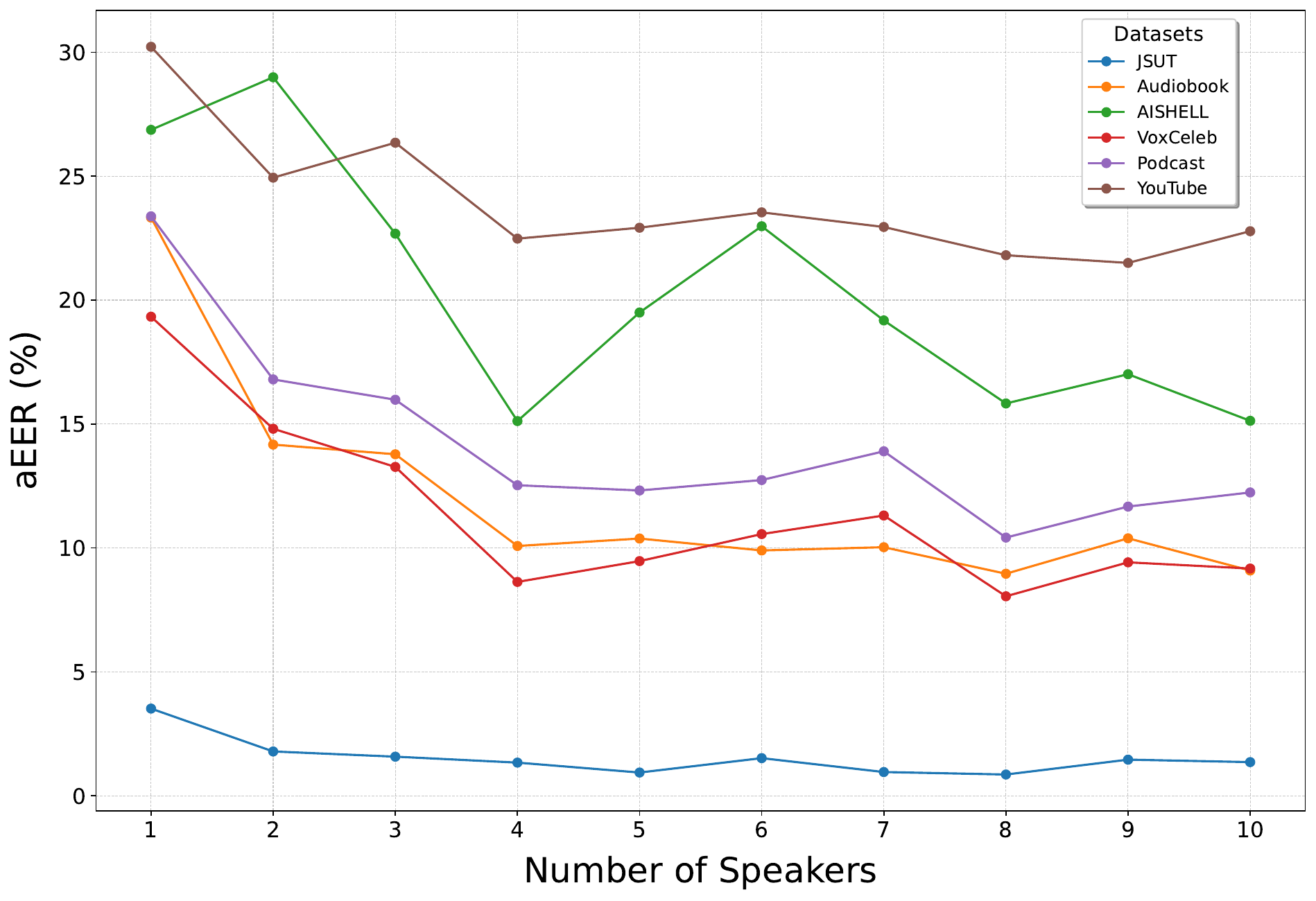}}
    \caption{Average EERs reported with models trained on increasing number of speakers with HFG generated audio samples (LibriTTS, train-clean-360). For each dataset average EER on all 12 test vocoders are plotted. Test samples are also generated using HFG vocoder. aEER drops significantly when the number of speakers is increased to four. No significant performance gain was observed thereafter, for most cases.}
     \label{fig:hfg_in-domain}
\end{center}
\vskip -0.2in 
\end{figure}

We observed that for each of the distribution shifts, EER\% decreased as number of speakers in training are increased from 1 to 4 (\autoref{fig:hfg_in-domain}). While performance on certain datasets like AISHELL exhibits inconsistent behavior, the aEER obtained on this dataset with 4 speakers in training is comparable to including ten speakers in training. These experimental results suggest that a dataset with four speakers may be sufficient for training, and increasing the number of speakers beyond this threshold might not provide substantial benefits to the model’s performance. 

\begin{figure}[H]
\vskip 0.2in 
\begin{center}
\centerline{\includegraphics[width=0.8\textwidth]{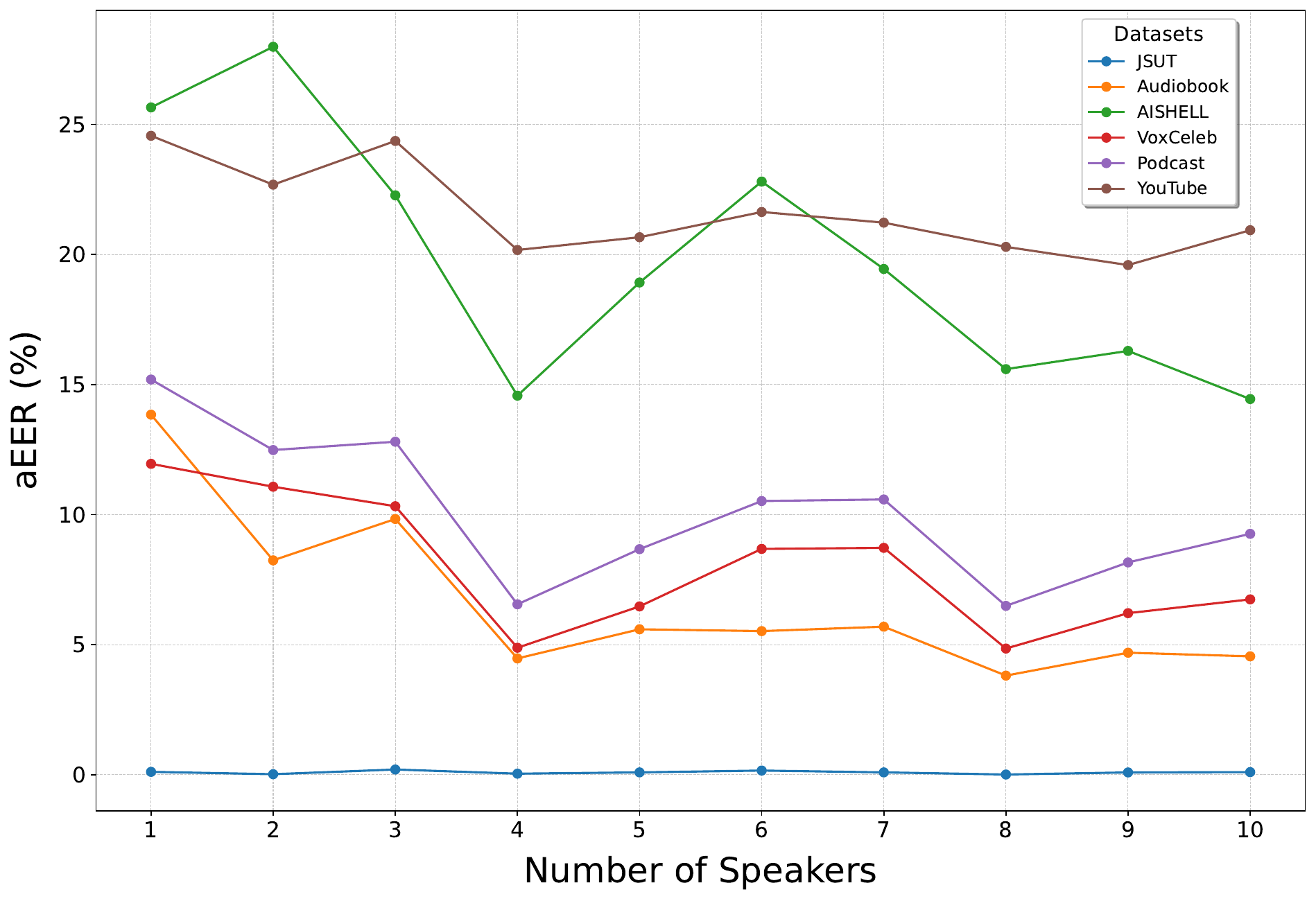}}
    \caption{Average EERs reported with models trained on increasing number of speakers with HFG generated audio samples (LibriTTS, train-clean-360). For each dataset average EER on all 12 test vocoders are plotted. Test samples are generated using PWG vocoder. aEER drops significantly when number of speakers are increased to four. No significant performance gain was observed thereafter, for most cases.}
    \label{fig:hfg_out-of-domain}
\end{center}
\vskip -0.2in 
\end{figure}

For JSUT dataset, aEER of model trained with one speaker is 3.52\% and it further decreases to 1.34\% with 4 speakers in training. While training with all 10 speakers results in slight increase in aEER to 1.36\%. Although training on 8 speakers led to a reduction in aEER of 11\% in case of audiobook dataset; 35\% in case of JSUT dataset; 16\% in case of Podcast dataset. Only 3\% and 7\% reduction was observed for YouTube and VoxCeleb dataset respectively. Moreover, we observed similar results when evaluation was done on an out-of-domain vocoder PWG (see \autoref{fig:hfg_out-of-domain}).
\paragraph{Discussion} These results suggest that there are quickly diminishing returns from including more than 4 speakers in training. This is a positive finding, since it suggests that fewer resources need to be spent identifying a large number of speakers. Detailed results can be found below.
\begin{table}[H]
\caption{Model trained on synthetic audio samples generated using HFG vocoder, train-clean360 LibriTTS and tested on HFG vocoder generated samples with JSUT as source dataset. No significant drop in EER \% was observed by increasing number of speakers beyond four }
\label{tab:inc-spk-jsut}
\vskip 0.15in
\begin{center}
\scriptsize
\resizebox{\textwidth}{!}{
\begin{tabular}{|c|c|c|c|c|c|c|c|c|c|c|}
\hline
Exp no. & \textbf{spk1}& \textbf{spks2} &\textbf{spks3}  &\textbf{spks4}  &\textbf{spks5}  &\textbf{spks6}  &\textbf{spks7}  &\textbf{spks8}   &\textbf{spks9}  &\textbf{spks10}\\
\hline 
1 &0.84&0.56&0.82&0.82&1.32&0.86&0.58&0.66&1.22 &1.02\\ 
\hline
2 &1.42&0.98&0.90&1.14&1.04&1.86&0.72&1.06&1.02 &1.28\\ 
\hline
3 &1.88&1.24&1.66&1.28&0.88&1.02&0.90&0.54&1.02 &1.96\\ 
\hline
4 &11.56&4.04&2.66&2.22&0.82&2.60&1.72&0.86&2.20 &1.32\\ 
\hline
5 &1.94&2.14&1.88&1.26&0.68&1.30&0.88&1.18&1.84 &1.24\\ 
\hline
\textbf{aEER} &\textcolor{red}{3.52} &1.79&1.58&1.34&0.94&1.52&0.96&\textcolor{darkgreen}{0.86}&1.46 &1.36\\ 
\bottomrule
\end{tabular}}%
\end{center}
\vskip -0.1in
\end{table}

\begin{table}[H]
\caption{Model trained on synthetic audio samples generated using HFG vocoder, train-clean360 LibriTTS and tested on HFG vocoder generated samples with Audiobook as source dataset. No significant drop in EER \% was observed by increasing number of speakers beyond four}
\label{tab:inc-spk-audiobook}
\vskip 0.15in
\begin{center}
\scriptsize
\resizebox{\textwidth}{!}{
\begin{tabular}{|c|c|c|c|c|c|c|c|c|c|c|}
\hline
Exp no. & \textbf{spk1}& \textbf{spks2} &\textbf{spks3}  &\textbf{spks4}  &\textbf{spks5}  &\textbf{spks6}  &\textbf{spks7}  &\textbf{spks8}   &\textbf{spks9}  &\textbf{spks10}\\
\hline 
1 &29.59&16.91&9.94&9.14&12.84&10.07&12.74&10.51&14.31 &9.11\\ 
\hline
2 &23.29&12.50&10.71&10.30&9.69&9.50&9.00&8.85&10.24 &8.62\\ 
\hline
3 &12.36&10.55&8.00&9.06&8.00&8.26&7.09&7.06&8.24 &9.29\\ 
\hline
4 &23.75&17.48&28.89&9.72&12.35&11.49&10.29&9.98&9.98 &9.31\\ 
\hline
5 &27.66&13.44&11.36&12.22&9.03&10.18&11.07&8.44&9.21 &9.12\\ 
\hline
\textbf{aEER} &\textcolor{red}{23.33}&14.17&13.78&10.08&10.38&9.90&10.03&\textcolor{darkgreen}{8.96} &10.39 & 9.09\\ 
\bottomrule
\end{tabular}}%
\end{center}
\vskip -0.1in
\end{table}

\begin{table}[H]
\caption{Model trained on synthetic audio samples generated using HFG vocoder, train-clean360 LibriTTS and tested on HFG vocoder generated samples with Podcast as source dataset. No significant drop in EER \% was observed by increasing number of speakers beyond four}
\label{tab:inc-spk-podcast}
\vskip 0.15in
\begin{center}
\scriptsize
\resizebox{\textwidth}{!}{
\begin{tabular}{|c|c|c|c|c|c|c|c|c|c|c|}
\hline
Exp no. & \textbf{spk1}& \textbf{spks2} &\textbf{spks3}  &\textbf{spks4}  &\textbf{spks5}  &\textbf{spks6}  &\textbf{spks7}  &\textbf{spks8}   &\textbf{spks9}  &\textbf{spks10}\\
\hline 
1 &33.29&20.36&11.33&11.61&13.78&12.40&21.08&11.85&12.42 &10.58\\ 
\hline
2 &23.01&15.50&11.09&12.48&12.38&12.55&9.79&8.71&12.54 &13.51\\ 
\hline
3 &14.47&15.04&12.79&9.90&10.11&8.88&10.86&7.31&9.17&12.29\\ 
\hline
4 &25.19&21.00&29.80&10.99&15.09&16.03&15.10&12.38&13.02 &10.36\\ 
\hline
5 &20.98&12.11&14.90&17.71&10.28&13.84&12.68&11.89&11.24 &14.49\\ 
\hline
\textbf{aEER} &\textcolor{red}{23.38}&16.80&15.98&12.53&12.32&12.74&13.90&\textcolor{darkgreen}{10.42}&11.67 &12.24\\ 
\bottomrule
\end{tabular}}%
\end{center}
\vskip -0.1in
\end{table}

\begin{table}[H]
\caption{Model trained on synthetic audio samples generated using HFG vocoder, train-clean360 LibriTTS and tested on HFG vocoder generated samples with Youtube as source dataset. No significant drop in EER \% was observed by increasing number of speakers beyond four}
\label{tab:in-spk-yt}
\scriptsize
\vskip 0.15in
\begin{center}
\resizebox{\textwidth}{!}{
\begin{tabular}{|c|c|c|c|c|c|c|c|c|c|c|}
\hline
Exp no. & \textbf{spk1}& \textbf{spks2} &\textbf{spks3}  &\textbf{spks4}  &\textbf{spks5}  &\textbf{spks6}  &\textbf{spks7}  &\textbf{spks8}   &\textbf{spks9}  &\textbf{spks10}\\
\hline 
1 &36.99&28.33&24.41&24.89&23.50&26.08&27.15&23.44&20.96 &22.99\\ 
\hline
2 &31.32&23.81&21.48&21.76&23.75&22.24&21.70&20.05&21.05 &21.45\\ 
\hline
3 &24.45&22.90&20.64&19.77&19.33&19.86&20.07&19.21&20.07 &22.44\\ 
\hline
4 &31.14&28.22&43.36&21.32&25.31&25.52&23.94&21.72&23.13 &24.66\\ 
\hline
5 &27.74&21.44&21.86&24.68&22.74&24.01&21.93&24.64&22.33 &22.37\\ 
\hline
\textbf{aEER} &\textcolor{red}{30.22}&24.94&26.35&22.48&22.92&23.54&22.95&21.81& \textcolor{darkgreen}{21.50}&22.78\\ 
\bottomrule
\end{tabular}}%
\end{center}
\vskip -0.1in
\end{table}

\begin{table}[H]
\caption{Model trained on synthetic audio samples generated using HFG vocoder, train-clean360 LibriTTS and tested on HFG vocoder generated samples with VoxCeleb as source dataset. No significant drop in EER \% was observed by increasing number of speakers beyond four}
\label{tab:inc-spk-voxceleb}
\scriptsize
\vskip 0.15in
\begin{center}
\resizebox{\textwidth}{!}{
\begin{tabular}{|c|c|c|c|c|c|c|c|c|c|c|}
\hline
Exp no. & \textbf{spk1}& \textbf{spks2} &\textbf{spks3}  &\textbf{spks4}  &\textbf{spks5}  &\textbf{spks6}  &\textbf{spks7}  &\textbf{spks8}   &\textbf{spks9}  &\textbf{spks10}\\
\hline 
1 &24.00&16.86&8.24&7.83&10.38&11.75&18.15&9.06&10.23&8.14\\ 
\hline
2 &23.40&17.07&8.71&7.91&8.86&8.69&7.44&6.52&7.63 &9.62\\ 
\hline
3 &12.98&13.25&9.29&7.79&7.32&6.70&7.71&6.23&7.14&11.12\\ 
\hline
4 &20.53&18.85&32.21&7.75&14.01&17.68&13.23&8.76&13.21&8.33\\ 
\hline
5 &15.77&8.06&7.91&11.88&6.81&8.02&10.03&9.72&8.90 &8.67\\ 
\hline
\textbf{aEER} &\textcolor{red}{19.33}&14.81&13.27&8.63&9.47&10.56&11.31&\textcolor{darkgreen}{8.05}&9.42 &9.17\\ 
\bottomrule
\end{tabular}}%
\end{center}
\vskip -0.1in
\end{table}

\begin{table}[H]
\caption{Model trained on synthetic audio samples generated using HFG vocoder, train-clean360 LibriTTS and tested on HFG vocoder generated samples with AISHELL as source dataset. No significant drop in EER \% was observed by increasing number of speakers beyond four}
\label{tab:inc-spks-aishell}
\vskip 0.15in
\begin{center}
\scriptsize
\resizebox{\textwidth}{!}{
\begin{tabular}{|c|c|c|c|c|c|c|c|c|c|c|}
\hline
Exp no. & \textbf{spk1}& \textbf{spks2} &\textbf{spks3}  &\textbf{spks4}  &\textbf{spks5}  &\textbf{spks6}  &\textbf{spks7}  &\textbf{spks8}   &\textbf{spks9}  &\textbf{spks10}\\
\hline 
1 &28.84&24.34&18.91&8.32&29.01&28.31&21.72&22.33&20.19 &15.51\\ 
\hline
2 &38.51&30.63&24.17&11.27&20.13&16.09&18.74&13.54&13.47 &16.90\\ 
\hline
3 &21.40&35.49&13.43&10.98&15.24&12.23&16.57&5.18&11.92 &17.22\\ 
\hline
4 &27.02&34.76&38.68&23.23&24.12&37.90&23.76&19.05&27.85 &11.92\\ 
\hline
5 &18.61&19.74&18.22&21.80&9.04&20.37&15.11&19.07&11.65 &14.11\\ 
\hline
\textbf{aEER} &26.87&\textcolor{red}{28.99}&22.68&\textcolor{darkgreen}{15.12}&19.50&22.98&19.18&15.83&17.01 &15.13\\ 
\bottomrule
\end{tabular}}%
\end{center}
\vskip -0.1in
\end{table}

\subsection{Performance on out-of-domain vocoder}
In the below experiments we investigate if similar behavior is observed with out-of-domain vocoder used in evaluations. Here, the out-of-domain vocoder used is PWG 
\begin{table}[H]
\caption{Model trained on synthetic audio samples generated using HFG vocoder, train-clean360 LibriTTS and tested on PWG vocoder generated samples with JSUT as source dataset.}
\label{tab:inc-spk-jsut-pwg}
\vskip 0.15in
\begin{center}
\scriptsize
\resizebox{\textwidth}{!}{
\begin{tabular}{|c|c|c|c|c|c|c|c|c|c|c|}
\hline
Exp no. & \textbf{spk1}& \textbf{spks2} &\textbf{spks3}  &\textbf{spks4}  &\textbf{spks5}  &\textbf{spks6}  &\textbf{spks7}  &\textbf{spks8}   &\textbf{spks9}  &\textbf{spks10}\\
\hline 
1 &0.02 & 0.02 & 0.08 & 0.0 & 0.10 & 0.02 & 0.40 & 0.0 & 0.0 & 0.0\\ 
\hline
2 &0.02& 0.04 & 0.0 & 0.0 & 0.18 & 0.16 & 0.04 & 0.0 & 0.0 & 0.22\\ 
\hline
3 &0.22 & 0.02 & 0.58 & 0.06 & 0.20 & 0.0& 0.02& 0.02 & 0.08 & 0.02\\ 
\hline
4 &0.30& 0.04 & 0.34 & 0.06 & 0.0 & 0.02 & 0.0& 0.0 & 0.02 & 0.0\\ 
\hline
5 &0.0 & 0.0 & 0.04 & 0.10 & 0.0 & 0.64 & 0.0 & 0.02 & 0.38 & 0.28\\ 
\hline
\textbf{aEER} & \textcolor{red}{0.11}& 0.02 & 0.20 & 0.04 & 0.09 & 0.16 & 0.09 & \textcolor{darkgreen}{0.008} & 0.088 & 0.10\\ 
\bottomrule
\end{tabular}}%
\end{center}
\vskip -0.1in
\end{table}

\begin{table}[H]
\caption{Model trained on synthetic audio samples generated using HFG vocoder, train-clean360 LibriTTS and tested on PWG vocoder generated samples with AISHELL as source dataset.}
\label{tab:inc-spk-aishell-pwg}
\vskip 0.15in
\begin{center}
\scriptsize
\resizebox{\textwidth}{!}{
\begin{tabular}{|c|c|c|c|c|c|c|c|c|c|c|}
\hline
Exp no. & \textbf{spk1}& \textbf{spks2} &\textbf{spks3}  &\textbf{spks4}  &\textbf{spks5}  &\textbf{spks6}  &\textbf{spks7}  &\textbf{spks8}   &\textbf{spks9}  &\textbf{spks10}\\
\hline 
1 &22.71 & 19.28 & 16.86 & 5.58 & 26.10 & 27.45 & 23.02 & 21.12 & 16.57 & 12.09\\ 
\hline
2 &38.51 & 30.63 & 24.17 & 11.27 & 20.13 & 16.09 &  18.74 & 13.54 & 13.47 & 16.90\\ 
\hline
3 &21.40 & 35.49 & 13.43 & 10.98 & 15.24 & 12.23 & 16.57 & 5.18 & 11.92 & 17.22\\ 
\hline
4 &27.02 & 34.76 & 38.68 & 23.23 & 24.12& 37.90 & 23.76 & 19.05 & 27.85 & 11.92\\ 
\hline
5 &18.61 & 19.74 & 18.22 & 21.80 & 9.04 & 20.37 & 15.11 & 19.07 & 11.65 & 14.11\\ 
\hline
\textbf{aEER} &25.65 & \textcolor{red}{27.98} & 22.27 & 14.57 & 18.92 & 22.80 & 19.44 & 15.59 & 16.29 & \textcolor{darkgreen}{14.44}\\ 
\bottomrule
\end{tabular}}%
\end{center}
\vskip -0.1in
\end{table}

\begin{table}[H]
\caption{Model trained on synthetic audio samples generated using HFG vocoder, train-clean360 LibriTTS and tested on PWG vocoder generated samples with VoxCeleb as source dataset.}
\label{tab:inc-spk-}
\vskip 0.15in
\begin{center}
\scriptsize
\resizebox{\textwidth}{!}{
\begin{tabular}{|c|c|c|c|c|c|c|c|c|c|c|}
\hline
Exp no. & \textbf{spk1}& \textbf{spks2} &\textbf{spks3}  &\textbf{spks4}  &\textbf{spks5}  &\textbf{spks6}  &\textbf{spks7}  &\textbf{spks8}   &\textbf{spks9}  &\textbf{spks10}\\
\hline 
1 &16.18 & 12.39 & 3.40 & 2.44 & 4.76& 9.72 & 17.37 & 4.61 & 4.45 & 4\\ 
\hline
2 &16.22 & 12.47 & 4.90 & 5.25 & 8.37 & 7.12 & 4.55 & 2.75 & 5.58 & 10.36\\ 
\hline
3 &10.29 & 11.44 & 6.54 & 2.95 & 3.44 & 4.06 & 4.88 & 2.40 & 3.87 & 7.42\\ 
\hline
4 &11.77 & 16.66 & 32.64 & 4.24 & 12.43 & 16.82 & 11.53 & 8.00 & 13.60 & 4.98\\ 
\hline
5 &5.33 & 2.40 & 4.16 & 9.52 & 3.38 & 5.68 & 5.29 & 6.50 & 3.59 & 6.34\\ 
\hline
\textbf{aEER} &\textcolor{red}{11.95} & 11.07 & 10.32 & 4.88 & 6.47 & 8.68 & 8.72 & \textcolor{darkgreen}{4.85} & 6.21 & 6.74\\ 
\bottomrule
\end{tabular}}%
\end{center}
\vskip -0.1in
\end{table}

\begin{table}[H]
\caption{Model trained on synthetic audio samples generated using HFG vocoder, train-clean360 LibriTTS and tested on PWG vocoder generated samples with Audiobook as source dataset.}
\label{tab:inc-spk-pwg-audiobook}
\vskip 0.15in
\begin{center}
\scriptsize
\resizebox{\textwidth}{!}{
\begin{tabular}{|c|c|c|c|c|c|c|c|c|c|c|}
\hline
Exp no. & \textbf{spk1}& \textbf{spks2} &\textbf{spks3}  &\textbf{spks4}  &\textbf{spks5}  &\textbf{spks6}  &\textbf{spks7}  &\textbf{spks8}   &\textbf{spks9}  &\textbf{spks10}\\
\hline 
1 &20.59 & 12.49 & 4.31 & 2.15 & 6.95 & 6.38 & 11.09 & 4.86 & 6.07 & 3.63\\ 
\hline
2 &14.83 & 7.70 & 6.21 & 5.19 & 7.14 & 5.38 & 4.24 & 2.40 & 5.40 & 6.81\\ 
\hline
3 &9.11 & 5.73 & 3.74 & 3.30 & 2.22 & 2.29 & 2.80 & 1.53 & 2.41 & 3.45\\ 
\hline
4 &10.46 & 11.98 & 28.52 & 3.71 & 8.80 & 7.62 & 6.79 & 6.34 & 6.57 & 2.69\\ 
\hline
5 &14.25 & 3.34 & 6.39 & 8.02 & 2.84 & 5.94 & 3.56 & 3.95 & 3.04 & 6.20\\ 
\hline
\textbf{aEER} &\textcolor{red}{13.84} & 8.24 & 9.83 & 4.47 & 5.59 & 5.52 & 5.69 & \textcolor{darkgreen}{3.81} & 4.69 & 4.55\\ 
\bottomrule
\end{tabular}}%
\end{center}
\vskip -0.1in
\end{table}

\begin{table}[H]
\caption{Model trained on synthetic audio samples generated using HFG vocoder, train-clean360 LibriTTS and tested on PWG vocoder generated samples with Podcast as source dataset.}
\label{tab:inc-spk-pwg-podcast}
\vskip 0.15in
\begin{center}
\scriptsize
\resizebox{\textwidth}{!}{
\begin{tabular}{|c|c|c|c|c|c|c|c|c|c|c|}
\hline
Exp no. & \textbf{spk1}& \textbf{spks2} &\textbf{spks3}  &\textbf{spks4}  &\textbf{spks5}  &\textbf{spks6}  &\textbf{spks7}  &\textbf{spks8}   &\textbf{spks9}  &\textbf{spks10}\\
\hline 
1 &26.42 & 18.61 & 7.37 & 4.77 & 8.09 & 9.69 & 20.09 & 6.48 & 6.00 & 6.85\\ 
\hline
2 &14.14 & 10.18 & 7.02 & 9.46 & 11.58 & 11.19 & 5.94 & 3.55 & 9.64 & 14.49\\ 
\hline
3 &13.03 & 12.72& 9.61 & 5.66 & 5.35 & 7.05 & 7.42 & 3.88 & 6.09 & 7.12\\ 
\hline
4 &12.97 & 16.00 & 30.00& 5.79 & 12.85 & 13.98 & 12.24 & 10.87 & 11.87 & 6.39\\ 
\hline
5 &9.42 & 4.89 & 10.03 & 13.66 & 5.49 & 10.70 & 7.23 & 7.69& 7.21 & 11.48\\ 
\hline
\textbf{aEER} &\textcolor{red}{15.19} & 12.48 & 12.80 & 6.55 & 8.67 & 10.52 & 10.58 & \textcolor{darkgreen}{6.49} & 8.16 & 9.26\\ 
\bottomrule
\end{tabular}}%
\end{center}
\vskip -0.1in
\end{table}

\begin{table}[H]
\caption{Model trained on synthetic audio samples generated using HFG vocoder, train-clean360 LibriTTS and tested on PWG vocoder generated samples with YouTube as source dataset.}
\label{tab:inc-spk-pwg-yt}
\vskip 0.15in
\begin{center}
\scriptsize
\resizebox{\textwidth}{!}{
\begin{tabular}{|c|c|c|c|c|c|c|c|c|c|c|}
\hline
Exp no. & \textbf{spk1}& \textbf{spks2} &\textbf{spks3}  &\textbf{spks4}  &\textbf{spks5}  &\textbf{spks6}  &\textbf{spks7}  &\textbf{spks8}   &\textbf{spks9}  &\textbf{spks10}\\
\hline 
1 &30.80 & 25.93 & 21.61 & 22.12 & 20.37 & 24.08 & 25.88 & 21.30 & 19.20 & 20.76 \\ 
\hline
2 &26.96 & 22.00 & 19.72 & 19.55 & 22.02 & 20.21 & 20.35 & 18.77 & 18.52 & 20.37\\ 
\hline
3 &21.55 & 21.00 & 17.58 & 16.50 & 16.78 & 18.04 & 17.54 & 17.35 & 18.33 & 20.60\\ 
\hline
4 &23.37 & 26.65 & 43.86 & 18.93 & 23.66 & 23.68 & 21.91 & 20.76 & 21.35 & 22.39\\ 
\hline
5 &20.16 & 17.82 & 19.05 & 23.76 & 20.48 & 22.16 & 20.44 & 23.27 & 20.56 & 20.57\\ 
\hline
\textbf{aEER} &24.56 & 22.68 & 24.36 & 20.17 & 20.66 & 21.63 & 21.22 & 20.29 & \textcolor{darkgreen}{19.59} & 20.93\\ 
\bottomrule
\end{tabular}}%
\end{center}
\vskip -0.1in
\end{table}

%%%%%%%%%%%%%%%%%%%%%%%%%%%%%%%%%%%%%%%%%%%%%%%%%%%%%%%%%%%%

\newpage

\end{document}